\documentclass[12pt]{article}

\usepackage{graphicx}

\def\ltwid{\mathrel{\raise.3ex\hbox{$<$\kern-.75em\lower1ex\hbox{$\sim$}}}}
\def\gtwid{\mathrel{\raise.3ex\hbox{$>$\kern-.75em\lower1ex\hbox{$\sim$}}}}

\def\square{\kern1pt\vbox{\hrule height 1.2pt\hbox{\vrule width 1.2pt\hskip 3pt
   \vbox{\vskip 6pt}\hskip 3pt\vrule width 0.6pt}\hrule height 0.6pt}\kern1pt}
\def\overleftrightarrow#1{\vbox{\ialign{##\crcr
     $\leftrightarrow$\crcr\noalign{\kern-1pt\nointerlineskip}
     $\hfil\displaystyle{#1}\hfil$\crcr}}}

\begin{document}

\begin{titlepage}

\begin{flushright}
ITP-UU-13/12, SPIN-13/08 \\ UFIFT-QG-13-03, CCTP-2013-08
\end{flushright}

\begin{center}
{\bf The Perils of Analytic Continuation}
\end{center}

\begin{center}
S. P. Miao$^{1a,4}$, P. J. Mora$^{2b}$, N. C. Tsamis$^{3c}$ and R. P.
Woodard$^{2d}$
\end{center}

\begin{center}
\it{$^{1}$ Institute for Theoretical Physics \& Spinoza Institute, Utrecht
University \\ Leuvenlaan 4, Postbus 80.195, 3508 TD Utrecht, NETHERLANDS}\\
\end{center}

\begin{center}
\it{$^{2}$ Department of Physics, University of Florida \\
Gainesville, FL 32611, UNITED STATES}
\end{center}

\begin{center}
\it{$^{3}$ Institute of Theoretical Physics \& Computational Physics \\
Department of Physics University of Crete \\
GR-710 03 Heraklion, HELLAS}
\end{center}

\begin{center}
\it{$^{4}$ Department of Physics, National Cheng Kung University \\
No.1, University Road, Tainan City 701, TAIWAN}
\end{center}

\begin{center}
ABSTRACT
\end{center}

A nice paper by Morrison \cite{IAM} demonstrates the recent
convergence of opinion that has taken place concerning the graviton
propagator on de Sitter background. We here discuss the few points
which remain under dispute. First, the inevitable decay of tachyonic
scalars really does result in their 2-point functions breaking de
Sitter invariance. This is obscured by analytic continuation
techniques which produce formal solutions to the propagator equation
that are not propagators. Second, Morrison's de Sitter invariant
solution for the spin two sector of the graviton propagator involves
derivatives of the scalar propagator at $M^2 = 0$, where it is not
meromorphic unless de Sitter breaking is permitted. Third, de Sitter
breaking does not require zero modes. Fourth, the ambiguity Morrison
claims in the equation for the spin two structure function is fixed
by requiring it to derive from a mode sum. Fifth, Morrison's spin
two sector is not ``physically equivalent'' to ours because their
coincidence limits differ. Finally, it is only the noninvariant
propagator that gets the time independence and scale invariance of
the tensor power spectrum correctly.

\begin{flushleft}
\end{flushleft}

\begin{flushleft}
$^{a}$ e-mail: spmiao5@mail.ncku.edu.tw \hspace{.1cm} , \hspace{.5cm}
$^{b}$ e-mail: pmora@phys.ufl.edu \\
$^{c}$ e-mail: tsamis@physics.uoc.gr \hspace{1cm} , \hspace{.5cm}
$^{d}$ e-mail: woodard@phys.ufl.edu
\end{flushleft}

\end{titlepage}

\section{Introduction}\label{intro}

The increasingly compelling evidence for primordial inflation
\cite{data} has transformed quantum field theory on de Sitter space
from an esoteric exercise in mathematical physics to the essential
framework for deriving the initial conditions of observational
cosmology \cite{books}. The best understood of these initial
conditions take the form of primordial scalar and tensor
perturbations \cite{perts}. It is important to understand that these
perturbations were not present before inflation; they formed through
a time-dependent process in which virtual particles were ripped out
of the vacuum by the accelerated expansion of spacetime, and they
have preserved a memory of conditions at the time they were formed.

The basic perturbations are tree order effects and their power
spectra can be expressed in terms of plane wave mode functions,
evaluated after they have experienced first horizon crossing
\cite{reviews}. Of course the laws of physics are governed by an
interacting quantum field theory so primordial perturbations must
affect one another, at some level, and they must also affect other
particles. Loop effects of this sort could be expressed in terms of
mode sums but it is simplest to recognize these mode sums as
propagators. Hence the interest in scalar and tensor propagators on
a nearly de Sitter background.

The intrinsically time-dependent process through which inflationary
perturbations are generated would seem to preclude de Sitter
invariant propagators for either the massless, minimally coupled
(MMC) scalar or for the graviton. This was recognized quite early
for the MMC scalar by exhibiting the time dependence of its
coincidence limit \cite{VFLS}, and a formal proof was given soon
afterwards \cite{AF}. However, opinions about the graviton
propagator have been divided between cosmologists --- who argue that
it must break de Sitter invariance because free gravitons in
transverse-traceless-spatial gauge obey the same equation as MMC
scalars \cite{Grishchuk} --- and mathematical physicists who argue
that this is a gauge artifact \cite{HMM,MTW4}.

Explicit constructions of the graviton propagator have produced
equivocal results. On the one hand, adding a de Sitter invariant
gauge fixing term to the action yields propagator equations for
which explicit de Sitter invariant solutions have been given
\cite{INVPROP}, except for an infinite set of discrete choices of
the two gauge fixing parameters for which infrared divergences
preclude a de Sitter invariant solution \cite{IAEM}. On the other
hand, why should {\it any} choice of arbitrary gauge fixing
parameters lead to de Sitter breaking? And it was early shown that
the solution with a noncovariant gauge fixing parameter \cite{TW1}
manifests de Sitter breaking even when the compensating gauge
transformation is added \cite{Kleppe}.\footnote{The same de Sitter
breaking occurs using the different field variables favored by
Kitamoto and Kitazawa \cite{Kit2}.} This has led to a curious state
of affairs in which every complete, dimensionally regulated graviton
loop correction \cite{TW2,MW1,KW1,PMW,SPM,LW} has been made using a
propagator that mediates plausible de Sitter breaking effects ---
for example, that scattering with inflationary gravitons induces
secular growth of fermion wave functions \cite{MW1} --- which
mathematical physicists suspect to be gauge artifacts.

Four recent insights have partially resolved this unsatisfactory
situation:
\begin{itemize}
\item{There is an obstacle to adding invariant gauge fixing terms on
any manifold, such as de Sitter, with a linearization instability
\cite{MTW1};\footnote{Ignoring this problem in scalar quantum
electrodynamics leads to on-shell singularities in the scalar
self-mass-squared \cite{KW2}.}}
\item{Power law infrared divergences are incorrectly subtracted off
by the analytic continuation techniques routinely employed by
mathematical physicists, leading to formal solutions of the
propagator equation which are not true propagators \cite{MTW2};}
\item{When de Sitter invariant gauges are imposed as strong operator
conditions --- as opposed to the average conditions effected by
adding gauge fixing functions --- the resulting propagators show de
Sitter breaking \cite{MTW3,KMW,PMTW1}; and}
\item{The old, noncovariant gauge propagator \cite{PMW}, and all of the
new covariant gauge ones \cite{PMTW2}, give a result that
mathematical physicists accept as correct for the linearized
Weyl-Weyl correlator \cite{Kouris}.\footnote{The mathematical
physics computation \cite{Kouris} had a number of significant errors
that were discovered by comparison with the cosmological result
\cite{PMW} and then corrected \cite{Atsuchi}.}}
\end{itemize}
The second point also explains the isolated infrared divergences
long encountered in constructions of the graviton propagator
\cite{IAEM}. Analytic continuation techniques only register {\it
logarithmic} divergences \cite{KK,JMPW}, and the problematic
parameter values are just those for which a power law infrared
divergence happens to become logarithmic.

A recent paper by Morrison \cite{IAM} reveals how close the two
sides have grown. In particular, he has exploited the formalism of
covariant projection operators acting on scalar structure functions
which was developed to represent the graviton self-energy \cite{SPW}
and later applied to the propagator in exact de Donder gauge
\cite{MTW3}. Using this formalism he has explained carefully what
must be done to extract a de Sitter invariant solution, and he has
exhibited the rather small differences in the structure functions
which distinguish the de Sitter breaking solution from the invariant
one. He has also demonstrated that the two propagators agree when
smeared with transverse-traceless test functions in the sense of
Fewster and Hunt \cite{FH}.

It would be reasonable to infer from Morrison's work that
mathematical physicists have no further objections to the
noncovariant gauge propagator \cite{TW1} which has been used for
every complete, dimensionally regulated graviton loop so far
computed \cite{TW2,MW1,KW1,PMW,SPM,LW}. Nevertheless, there are a
few issues that remain controversial. These concern the validity of
the analytic continuation techniques used in Morrison's construction
and the distinction between formal solutions of the propagator
equation and true propagators.

This paper contains five sections of which the first is this
Introduction. In section~\ref{tachyon} we consider the contention by
mathematical physicists \cite{Higuchi,IAM} that minimally coupled
scalars with tachyonic masses $M^2 < 0$ nonetheless possess de
Sitter invariant propagators on $D$-dimensional de Sitter space with
Hubble constant $H$, except for the discrete values $M^2 = -N (N +
D-1) H^2$, where $N = 0, 1, 2, \dots$ In section~\ref{real} we 
re-visit classic work on the MMC scalar \cite{AF,BA} to debunk the 
more recent argument that its de Sitter breaking derives from the 
isolated zero mode in global coordinates, which gravitons lack. In 
fact the MMC scalar's infrared finite de Sitter breaking derives 
from the late time approach to scale invariance and time 
independence of its power spectrum in the ultraviolet, not the 
infrared. Both of these features are shared by the graviton. The 
infrared divergences of open coordinates derive from an 
infinite number of modes being in this saturated state at 
finite times. Section~\ref{spin2} discusses what is wrong with the 
construction which leads to a de Sitter invariant spin two structure 
function, why the coincidence limit \cite{KMW} shows that the de 
Sitter invariant solution \cite{IAM} is not physically equivalent to 
the de Sitter breaking one \cite{MTW3}, and why the time 
independence and scale invariance of the tensor power spectrum imply 
that the de Sitter breaking solution is correct. Minor points are 
that the coincidence limit of the graviton propagator appears in 
even simple one loop diagrams \cite{TW2,MW1,KW1,PMW,SPM,LW} --- so 
it cannot be dismissed --- and that no sequence of the 
transverse-traceless test functions of Fewster and Hunt \cite{FH} 
approaches a delta function --- so they are not analogous to the 
scalar smearing functions long employed by mathematical physicists. 
Our conclusions are summarized in section~\ref{discuss}.

\section{Scalar Tachyons}\label{tachyon}

Some mathematical physicists believe strongly that tachyonic scalars
possess de Sitter invariant propagators for any mass-squared which
avoids the discrete values $M^2 = -N(N + D-1) H^2$, where $N$ is a 
nonnegative integer \cite{Higuchi,IAM}. They believe this because, 
except for those special masses, the scalar propagator equation,
\begin{equation}
\sqrt{-g} \Bigl[ \square \!-\! M^2\Bigr] i\Delta(x;x') \equiv
\Bigl[\partial_{\mu} \Bigl(\sqrt{-g} g^{\mu\nu} \partial_{\nu}\Bigr)
\!-\! M^2 \sqrt{-g} \Bigr] i\Delta(x;x') = i \delta^D(x \!-\! x') 
\; , \label{propeqn}
\end{equation}
has a de Sitter invariant solution,
\begin{equation}
i\Delta^{\rm dS}(x;x') = \frac{H^{D-2}}{(4 \pi)^{\frac{D}2}}
\frac{\Gamma(\frac{D-1}2 \!+\! \nu) \Gamma(\frac{D-1}2 \!-\!
\nu)}{\Gamma(\frac{D}2)} \, \mbox{}_2 F_1\Bigl( \frac{D\!-\!1}2
\!+\! \nu, \frac{D\!-\!1}2 \!-\! \nu ; \frac{D}2 ; 1 \!-\!
\frac{y}4\Bigr) \; , \label{mathsol}
\end{equation}
where the index $\nu$ and the de Sitter length function $y(x;x')$
depend the upon $M^2$ and the invariant length $\ell(x;x')$ as,
\begin{equation}
\nu \equiv \sqrt{ \Bigl(\frac{D\!-\! 1}2\Bigr)^2 \!-\!
\frac{M^2}{H^2} } \qquad , \qquad y(x;x') \equiv 4 \sin^2\Bigl(
\frac12 H \ell(x;x')\Bigr) \; . \label{moremath}
\end{equation}
This belief is perplexing to cosmologists who feel that tachyonic
particles must decay, even on de Sitter space, and that this decay
is an inherently time-dependent process which depends upon when the
state is released and hence must break de Sitter invariance because
the result does not depend only on the observation time. Much of the
observed phenomenology of the Standard Model would not make sense
otherwise. In this section we first explain, in very simple terms,
why tachyonic scalars decay, and then how the use of analytic
continuation techniques can lead to employing formal solutions to
the scalar propagator equation which are not true propagators. The
key point is that the quantum mechanical requirement that states 
have finite, positive norm imposes restrictions on analytic
continuation which are being violated to discard the de Sitter
breaking secular terms. The section closes with a discussion of the 
unacceptable phenomenology implied by this practice.

\subsection{Why tachyonic scalars decay}

The first thing to understand is that tachyonic scalars decay
equally in open coordinates,
\begin{equation}
ds^2 = -dt^2 + e^{2 H t} d\vec{x} \!\cdot\! d\vec{x} \; ,
\label{open}
\end{equation}
and in closed coordinates (which we specialize to $D=4$),
\begin{equation}
ds^2 = -d\tau^2 + H^{-2} \cosh^2(H \tau) \Bigl[ d\chi^2 +
\sin^2(\chi) d\theta^2 + \sin^2(\chi) \sin^2(\theta) d\phi^2\Bigr]
\; . \label{closed}
\end{equation}
Another important point is that the decay occurs for each mode
separately, so one need never worry about more than a single degree
of freedom evolving in whatever is the appropriate time. This single
degree of freedom is known as a mode. The natural modes for open
coordinates are spatial plane waves $e^{\vec{k} \cdot \vec{x}}$ and
the associated mode functions are $u(t,k)$; for closed coordinates
the natural modes are the 4-dimensional spherical harmonics
$Y_{k\ell m}(\chi,\theta,\phi)$ and the associated mode functions
are $u_{k}(\tau)$. Finally, it is important to understand that there
is only one scalar field operator, $\varphi(x)$, and it can be
expanded in either coordinate system,
\begin{eqnarray}
\varphi(x) & \!\!\! = \!\!\! & \int \!\! \frac{d^3k}{(2\pi)^3}
\Biggl\{ u(t,k) e^{i \vec{k} \cdot \vec{x}} a(\vec{k}) \!+\!
u^*(t,k) e^{-i \vec{k} \cdot \vec{x}} a^{\dagger}(\vec{k}) \Biggr\}
\; , \qquad \label{openfree} \\
& \!\!\! = \!\!\! & \sum_{k=0}^{\infty} \sum_{\ell = 0}^{k} \sum_{m
= -\ell}^{\ell} \Biggl\{ u_{k}(\tau) Y_{k\ell m}(\chi,\theta,\phi)
a_{k\ell m} + u^*_{k}(\tau) Y^*_{k\ell m}(\chi,\theta,\phi)
a^{\dagger}_{k\ell m} \Biggr\} \; . \qquad \label{closedfree}
\end{eqnarray}

The massive scalar Lagrangian is,
\begin{equation}
\mathcal{L} = -\frac12 \partial_{\mu} \varphi \partial_{\nu} \varphi
g^{\mu\nu} \sqrt{-g} -\frac12 M^2 \varphi^2 \sqrt{-g} \; .
\label{Lagrangian}
\end{equation}
The Euler-Lagrange equation derived from (\ref{Lagrangian}) implies
the equations obeyed by $u(t,k)$ and $u_{k}(\tau)$,
\begin{eqnarray}
\Bigl[ \partial^2_t + 3 H \partial_t + k^2 e^{-2Ht} +
M^2\Bigr] u(t,k) & = & 0 \; , \label{openeqn} \\
\Bigl[ \partial^2_{\tau} + 3 H \tanh(H \tau) \partial_{\tau} + H^2 k
(k \!+\! 2) {\rm sech}^2(H \tau) + M^2\Bigr] u_{k}(\tau) & = & 0 \;
. \label{closedeqn}
\end{eqnarray}
Similarly, the canonical commutation relations (which ensure positive
norm states) derived from (\ref{Lagrangian}) fix the normalizations 
of the respective Wronskians,
\begin{eqnarray}
u(t,k) \dot{u}^*(t,k) \!-\! \dot{u}(t,k) u^*(t,k) & = & i \, e^{-3 H
t} \; , \label{openWron} \\
u_{k}(\tau) u^{\prime *}_{k}(\tau) \!-\! u'_{k}(\tau) u^*_{k}(\tau)
& = & i \, {\rm sech}^3(H \tau) \; . \label{closedWron}
\end{eqnarray}
The close relation between the open coordinate mode equations
(\ref{openeqn}) and (\ref{openWron}) and their closed coordinate
analogs (\ref{closedeqn}) and (\ref{closedWron}) is evident. This is
why claims for de Sitter invariance are no better justified
in closed coordinates than in open coordinates. We shall have more
to say about this in subsection~\ref{distinguishing} and
section~\ref{real}.

Equations (\ref{openeqn}) and (\ref{closedeqn}) make it quite
apparent both why tachyonic scalars decay, and why the decay is much
stronger on de Sitter than it is in flat space. Each equation
contains three terms which we can identify as an ``acceleration'', a
``friction force'' and a ``restoring force''. For open coordinates
these three terms are,
\begin{eqnarray}
{\rm Acceleration} & \longleftrightarrow & \ddot{u}(t,k) \; , \\
{\rm Friction\ Force} & \longleftrightarrow & -3 H \dot{u}(t,k) \; , \\
{\rm Restoring\ Force} & \longleftrightarrow & -\Bigl(M^2 \!+\! k^2
e^{-2Ht} \Bigr) u(t,k) \; .
\end{eqnarray}
For $M^2 > 0$ the restoring force makes the mode accelerate in the
direction opposite to its current value. So if $u(t,k)$ is positive,
its acceleration is negative; while if $u(t,k)$ is negative, its
acceleration is positive. The behavior in that case is either
under-damped or over-damped oscillations, depending upon the
relation between the friction force and the restoring force.

Tachyonic scalars have $M^2 < 0$, which tends to make the mode
accelerate in the same direction it already is: if $u(t,k)$ is
positive, a tachyonic mass term tends to make it accelerate in the
positive direction; and if $u(t,k)$ is negative, a tachyonic mass
term makes it accelerate in the negative direction. This effect is
resisted by the spatial gradient term $k^2 e^{-2Ht}$, but that
redshifts to zero at late times, so that all modes eventually
experience an anti-restoring force. That is why the instability is
worse than in flat space. The effect of a tachyonic mass is to make
the mode function $u(t,k) \rightarrow K(k) \times e^{\lambda t}$ grow 
exponentially at late times with time constant,
\begin{equation}
\lambda = -\frac32 H + \sqrt{\Bigl(\frac32 H\Bigr)^2 \!-\! M^2} \; .
\label{timeconst}
\end{equation}
The asymptotic late time behavior of the closed coordinate mode
function $u_{k}(\tau)$ is not a bit different for $\tau \rightarrow
+\infty$, but it also shows exponential growth for all modes as
$\tau \rightarrow -\infty$.

There should be nothing surprising about this analysis; it is
elementary Newtonian Mechanics. We are describing the modes in terms
of point particle motions and it is a fact that balls in uniform
gravitational fields tend to roll down parabolic hills with
ever-increasing speed. The Hubble friction term decreases the
exponential time constant somewhat but it cannot prevent the decay.

There {\it are} classical configurations which fail to experience
exponential growth for $M^2 < 0$. For example, if one chooses the
initial conditions so that $u(0,k) = 0$ and $\dot{u}(0,k) = 0$ then
the solution is $u(t,k) = 0$ for all time. One can also arrange the
initial conditions so that the mode approaches zero at late times.
If we were doing classical mechanics --- {\it or just trying to 
solve the propagator equation} --- there would be no objection to 
these solutions. However, the quantum mechanical requirement of 
positive norm states (\ref{openWron}-\ref{closedWron}) precludes the
mode function having just the exponentially falling solution.

A true propagator is the expectation value in some normalized state
we might call $\vert \Omega \rangle$ of the time-order product of
two field operators. Substituting the open coordinate and closed
coordinate free field expansions (\ref{openfree}) and
(\ref{closedfree}) gives,\footnote{Note that these expansions assume
the Heisenberg field equation is obeyed in the strong operator sense 
which is standard for quantum field theory. There have been attempts 
to avoid de Sitter breaking for tachyons by weakening the sense in
which the field equations hold \cite{BEM}.}
\begin{eqnarray}
\lefteqn{ \Bigl\langle \Omega \Bigl\vert T\Bigl[ \varphi(x)
\varphi(x') \Bigr] \Bigr\vert \Omega \Bigr\rangle } \nonumber \\
& & = \int \!\! \frac{d^3k}{(2\pi)^3} e^{i \vec{k} \cdot (\vec{x} -
\vec{x}')} \Biggl\{ \theta(t \!-\! t') u(t,k) u^*(t',k) \!+\!
\theta(t' \!-\! t) u^*(t,k) u(t',k) \Biggr\} , \qquad
\label{opensum} \\
& & = \sum_{k=0}^{\infty} \sum_{\ell = 0}^{k} \sum_{m=-\ell}^{\ell}
Y_{k \ell m} Y^*_{k\ell m} \Biggl\{ \theta(\tau \!-\! \tau')
u_k(\tau) u^*_k(\tau') \!+\! \theta(\tau' \!-\! \tau) u^*_k(\tau)
u_k(\tau') \Biggr\} . \qquad \label{closedsum}
\end{eqnarray}
Based on the analysis we have given of the mode equations
(\ref{openeqn}) and (\ref{closedeqn}), these mode sums cannot be de
Sitter invariant because the mode functions all show the same
exponential growth at late times,
\begin{eqnarray}
u(t,k) u^*(t',k) & \longrightarrow & \vert K(k)\vert^2 \times
e^{\lambda (t + t')} \; , \label{lateopen} \\
u_k(\tau) u^*_k(\tau') & \longrightarrow & \vert K_k \vert^2 \times
e^{\lambda (\tau + \tau')} \; . \label{lateclosed}
\end{eqnarray}
The positive-definite constant of proportionality depends upon the 
time at which the tachyonic mass term $M^2 < 0$ dominates the 
gradient term ($k^2 e^{-2Ht}$ for open coordinates, $H^2 k (k + 2) 
{\rm sech}^2(H \tau)$ for closed coordinates). For $k$ so small 
that the tachyonic mass term always dominates, the constant of 
proportionality depends upon the time at which the state was 
released.\footnote{The massless limit is also interesting. In that 
case all modes approach a constant ($H/\sqrt{2 k^3}$) at late 
times, which also precludes a de Sitter invariant result.}

At this stage one might wonder how anyone can extract a de Sitter
invariant result (\ref{mathsol}-\ref{moremath}) from mode sums
(\ref{opensum}) and (\ref{closedsum}) which must obviously break de
Sitter invariance for $M^2 < 0$. This is especially curious if, at
the same time, they admit there is de Sitter breaking for the
special values of $M^2 = -N(N + D-1) H^2$, where $N$ is a nonnegative
integer \cite{Higuchi,IAM}. This implies believing that, while a 
tachyonic scalar at one of those values decays, making the mass a 
little {\it more} tachyonic would prevent the decay!

We will see that the truth is less paradoxical. They have
analytically continued the mode sums (\ref{opensum}-\ref{closedsum})
from positive mass-squared --- for which there is no de Sitter
breaking decay --- to negative mass-squared. These analytic 
continuations preserve the equation of motion 
(\ref{openeqn}-\ref{closedeqn}) but not the canonical normalization
(\ref{openWron}-\ref{closedWron}) which ensures that states have
positive norm. The resulting mode sum is a formal solution to the 
propagator equation (\ref{propeqn}) which is not a true propagator 
in the sense of being the expectation value of the time-ordered 
product of two fields in the presence of some positive norm state 
\cite{TW3}. 

The problematic nature of analytic continuations which avoid de 
Sitter breaking is the same for both open and closed coordinates.
In both cases de Sitter breaking derives from more and more large 
$k$ modes reaching the saturated condition 
(\ref{lateopen}-\ref{lateclosed}) as time progresses, and this can
only be avoided by including negative norm states in the mode sum. 
However, there are interesting differences between the two
coordinate systems as regards infrared (small $k$) modes. In open
coordinates an infinite number of infrared modes are in the 
saturated condition (\ref{lateopen}) even at early times, whereas
only a finite number of closed coordinate modes have reached the
analogous form (\ref{lateclosed}) at any finite time. This results
in the open coordinate mode sum possessing an infrared divergence
which is absent from the closed coordinate mode sum. We will 
explore this infrared divergence further in the next two
subsections to see that the special thing about $M^2 = -N(N + D - 1) 
H^2$ is one of the power law infrared divergences happens to become
logarithmic, and hence visible to analytic regularization 
techniques. What happens in closed coordinates will be discussed in
section \ref{real}.

\subsection{Analytic continuations miss power law divergences}

In quantum field theory we are familiar with two sorts of
divergences: ultraviolet and infrared \cite{Weinberg}. Both require
regularization in order to render potentially divergent expressions
well-defined so that they can be analyzed. And it can happen that a
regularization technique fails to register the presence of a certain
class of divergences. For example, dimensionally regulating
\cite{dimreg} the quarticly divergent vacuum energy of free photons
gives zero,
\begin{equation}
2 \! \times \!\! \int \!\! \frac{d^3k}{(2\pi)^3} \, \frac{k}{2}
\qquad \longrightarrow \qquad (D \!-\! 2) \! \times \!\! \int \!\!
\frac{d^{D-1}k}{(2\pi)^{D-1}} \, \frac{k}{2} = 0 \; . \label{auto1}
\end{equation}
String theorists are familiar with how the application of zeta
function regularization \cite{zeta} to the central charge of the
Virasoro algebra converts the sum of positive integers into a
finite, negative number,
\begin{equation}
1 + 2 + 3 + \dots = \sum_{n=1}^{\infty} \frac1{n^{-1}} \qquad
\longrightarrow \qquad \lim_{s \rightarrow -1} \zeta(s) =
-\frac1{12} \; . \label{auto2}
\end{equation}

The suppression of obviously positive divergences in expressions
(\ref{auto1}) and (\ref{auto2}) is known as an ``automatic
subtraction''. The origin of the name can be understood if one
regulates the ill-defined sum on the left-hand side of (\ref{auto2})
using an exponential cutoff,
\begin{equation}
f(\epsilon) \equiv \sum_{n=1}^{\infty} n \times e^{-\epsilon n} =
\frac{e^{-\epsilon}}{(1 \!-\! e^{-\epsilon})^2} = \frac1{4
\sinh^2(\frac{\epsilon}2)} \; . \label{expcut1}
\end{equation}
This sort of method is known as a ``physical regularization''
because it shows the quadratic divergence in the unregulated limit
of $\epsilon \rightarrow 0$ ,
\begin{equation}
f(\epsilon) = \frac1{\epsilon^2} - \frac1{12} + O(\epsilon^2) \; .
\label{expcut2}
\end{equation}
The zeta function result (\ref{auto2}) is the $\epsilon \rightarrow
0$ limit of $f(\epsilon)$ with the power law divergence
$1/\epsilon^2$ subtracted off. Hence the name, ``automatic
subtraction''. Dimensional regularization can be similarly derived
by automatically subtracting power law divergences from a nonlocal
exponential cutoff which produces incomplete Gamma functions
\cite{GKW}.

Dimensional regularization and zeta function regularization are
known as ``analytic regularizations'' because they work by
considering the divergent expression to be an analytic function of
some parameter --- the spacetime dimension $D$ or the power $s$ to
which the eigenvalues of some operator are raised. The function
makes sense for certain parameter values and is then defined for all
others by analytic continuation. A hallmark of analytic continuation
techniques is that they fail to register power law divergences such
as those in expressions (\ref{auto1}) and (\ref{auto2}). On the
other hand, they do show the presence of logarithmic divergences.
One can see this by comparing the dimensionally regulated result for
the vacuum energy of a massive scalar,
\begin{eqnarray}
\lefteqn{\int \!\! \frac{d^3k}{(2\pi)^3} \, \frac12 \sqrt{k^2 + m^2}
\qquad \longrightarrow \qquad \mu^{4 - D} \!\! \int \!\!
\frac{d^{D-1}k}{(2\pi)^{D-1}} \, \frac12 \sqrt{k^2 + m^2} \; , } \\
& & \hspace{3cm} = -\frac{\Gamma(-\frac{D}2)}{2 (4\pi)^{\frac{D}2}}
\Bigl( \frac{\mu}{m}\Bigr)^{4-D} m^4 \; , \qquad \\
& & \hspace{3cm} = -\frac{m^4}{32 \pi^2} \Biggl[ \frac1{4 \!-\! D}
\!-\! \frac{\gamma}{2} \!+\! \frac34 \!+\! \frac12 \ln\Bigl(
\frac{4\pi \mu^2}{m^2}\Bigr) \!+\! O(4 \!-\! D)\Biggr] \; . \qquad
\label{Omegadim}
\end{eqnarray}
with the same quantity evaluated using a physical regularization
such as a momentum cutoff,
\begin{eqnarray}
\lefteqn{\int \!\! \frac{d^3k}{(2\pi)^3} \, \frac12 \sqrt{k^2 + m^2}
\qquad \longrightarrow \qquad \frac1{4\pi^2} \int_0^{\Lambda} \!\!
dk \, k^2 \sqrt{k^2 \!+\! m^2} \; , } \\
& & \hspace{2cm} = \frac1{32 \pi^2} \Biggl[ (2 \Lambda^3 \!+\! m^2
\Lambda) \sqrt{\Lambda^2 \!+\! m^2} \!-\! m^4 \ln\Biggl(
\frac{\Lambda}{m} \!+\! \sqrt{\frac{\Lambda^2}{m^2} \!+\! 1} \,
\Biggr)\Biggr] , \qquad \\
& & \hspace{2cm} = \frac1{32\pi^2} \Biggl[ 2 \Lambda^4 \!+\! 2 m^2
\Lambda^2 \!-\! m^4 \ln\Bigl( \frac{2\Lambda}{m}\Bigr) \!+\! \frac14
m^4 \!+\! O\Bigl(\frac{m^2}{\Lambda^2}\Bigr) \Biggr] . \qquad
\label{Omegacut}
\end{eqnarray}
The dimensionally regulated result (\ref{Omegadim}) agrees with the
logarithmic divergence of the momentum cutoff (\ref{Omegacut}) under
the correspondence,
\begin{equation}
\ln\Bigl( \frac{2 \Lambda}{m}\Bigr) \qquad \longleftrightarrow
\qquad \frac1{4 \!-\! D} \; .
\end{equation}
However, the quartic and quadratic divergences in the physical
regularization (\ref{Omegacut}) have been automatically subtracted
from the analytic regularization (\ref{Omegadim}).

\subsection{Infrared divergences must not be
subtracted}\label{mustnot}

Many features of infrared divergences are the same as for
ultraviolet divergences. In particular, both require regularization
for careful analysis, and analytic regularization techniques
automatically subtract off power law divergences from both. However,
what this means differs greatly.

As the name indicates, ultraviolet divergences originate from short
distance dynamics, far beyond the reach of any experiment. The cure
for an ultraviolet divergence is renormalization. In the sense of
Bogoliubov-Parasiuk-Hepp-Zimmerman \cite{BPHZ} (BPHZ) which is
relevant to quantum gravity, this means systematically adding new
local interactions to absorb primitive divergences, order-by-order
as they occur in the loop expansion. This can always be done, so the
only effect of an unphysical regularization technique which happens
to automatically subtract off a certain class of divergences is to
spare one the effort of constructing the relevant counterterms and
using them to absorb the subtracted divergences. In that case no
physical error results from using an analytic regularization
technique; indeed, it is the faster and simpler way to calculate.

Infrared divergences are very different. They come from dynamical
laws that have been thoroughly tested and are not subject to change.
The appearance of an infrared divergence is a quantum field theory's
way of indicating that something is physically wrong with the
question that is being asked of it. The proper course of action in
this case is to consider carefully what unphysical assumption led to
the divergence, and then pose a more physically meaningful question.
{\it It is a mistake to ignore the problem and continue to ask the
same, unphysical question.} However, the naive use of an analytic
regularization technique makes it difficult to recognize this
mistake, unless the infrared divergence happens to be logarithmic.

Flat space quantum electrodynamics (QED) provides the classic
example. Infrared divergences in that theory derive from the
exchange of soft photons between the external legs of exclusive
amplitudes. The physical problem with exclusive amplitudes is that
any real detector has a finite energy resolution, so there is no
experimental way to distinguish final states which differ by the
inclusion of a very low energy photon \cite{Weinberg}. When the
question being asked is made physically meaningful by including the
emission of arbitrary soft photons whose total energy is less than
some fixed detector resolution, the result becomes infrared finite
\cite{BN}. It also depends upon the detector resolution in a way
that agrees with experiment.

Quantum gravity with zero cosmological constant manifests very
similar infrared divergences, whose resolution is also achieved by
accounting for soft graviton emission \cite{SW2}. On the other hand,
infrared divergences can indicate the breakdown of unphysical
assumptions about the symmetry of a state. Veneziano considered the
perturbative massless limit of a real scalar field theory with cubic
self-interactions in flat space \cite{Veneziano}. He showed that
including the emission of soft scalars will not cure that theory's
infrared divergences. One must instead allow the vacuum to decay,
which of course breaks time translation invariance \cite{TW4}.

Ford and Parker discovered an example of great relevance to our
discussion in 1977 \cite{FP}. They considered a massless, minimally
coupled scalar on a spatially flat, FRW background,
\begin{equation}
ds^2 = -dt^2 + a^2(t) d\vec{x} \!\cdot\! d\vec{x} \qquad
\Longrightarrow \qquad H(t) \equiv \frac{\dot{a}}{a} \;\; , \;\;
\epsilon(t) \equiv -\frac{\dot{H}}{H^2} \; .
\end{equation}
The Fourier plane wave modes of this system are harmonic oscillators
with a time dependent mass ($m(t) \sim a^3(t)$) and frequency
$\omega(t) = k/a(t)$). At any instant the ground state energy of
each mode is $k/2a(t)$, although which state is the instantaneous
ground state changes with time. Ford and Parker specialized to power
law scale factors for which the slow roll parameter $\epsilon$ is an
arbitrary constant. A natural vacuum state --- and the one analogous
to Bunch-Davies vacuum for de Sitter \cite{BD} --- is the state
which was minimum energy in the distant past. Ford and Parker worked
in $D=4$ spacetime dimensions but it is useful to give the mode
function for general $D$,
\begin{equation} u(t,k) = a^{-(\frac{D-1}2)}
\sqrt{\frac{\pi}{4 (1 \!-\! \epsilon) H}} \, H^{(1)}_{\nu}\Bigl(
\frac{-k}{(1 \!-\! \epsilon) H a}\Bigr) \qquad {\rm with} \qquad \nu
= \frac{D \!-\! 1 \!-\! \epsilon}{2 (1 \!-\! \epsilon)} \; .
\label{epsmodes}
\end{equation}
The Fourier mode sum for the corresponding propagator would
be,\footnote{We might note that setting $\epsilon = 0$ and $\nu =
\sqrt{(\frac{D-1}2)^2 - \frac{M^2}{H^2}}$ in expression
(\ref{epsmodes}) gives the mode function for a massive scalar in
open de Sitter coordinates. The closed coordinate mode functions
\cite{MM} were derived by analytically continuing from Euclidean de
Sitter \cite{AH}.}
\begin{equation}
\int \!\! \frac{d^{D-1}k}{(2\pi)^{D-1}} \, e^{i \vec{k} \cdot
(\vec{x} - \vec{x}')} \Biggl\{ \theta(t \!-\! t') u(t,k) u^*(t',k)
\!+\! \theta(t' \!-\! t) u^*(t,k) u(t',k) \Biggr\} \; .
\label{naive}
\end{equation}
Ford and Parker showed that expression (\ref{naive}) suffers from
infrared divergences throughout the range from $\epsilon = 0$ (de
Sitter) to $\epsilon = \frac32$ (matter domination). This is obvious
from the small $k$ form of the mode functions (\ref{epsmodes}),
\begin{equation}
u(t,k) u^*(t',k) \longrightarrow \frac{4^{|\nu|} (1 \!-\!
\epsilon)^{|2\nu|} \Gamma^2(|\nu|)}{4 \pi (1 \!-\! \epsilon) \sqrt{H
a^{D-1} H' {a'}^{D-1}} } \times \Biggl[ \frac{ H a H'
a'}{k^2}\Biggr]^{|\nu|} \Biggl\{1 + O(k^2) \Biggr\} \; .
\label{smallk}
\end{equation}

For most values of $\epsilon$ and $D$ the infrared divergences Ford
and Parker found are of the power law type that would be invisible
to an analytic regularization. One can see this by evaluating
(\ref{naive}) for the infrared-finite case of $\epsilon > 2(D-1)/D$,
and noting that it produces an analytic function \cite{BD2,JP1},
\begin{equation}
\Bigl[(1 \!-\! \epsilon)^2 H H'\Bigr]^{\frac{D}2 -1}
\frac{\Gamma(\frac{D-1}2 \!+\! \nu) \Gamma(\frac{D-1}2 \!-\! \nu)}{
(4\pi)^{\frac{D}2} \Gamma(\frac{D}2)} \, \mbox{}_2
F_1\Bigl(\frac{D\!-\!1}2 \!+\! \nu, \frac{D-1}2 \!-\! \nu ;
\frac{D}2 ;1 \!-\! \frac{y}4\Bigr) \; , \label{dumb}
\end{equation}
where the constant $\epsilon$ length function (with infinitesimal
$\delta$ to fix the branch) is,
\begin{equation}
y(x;x') \equiv H a H' a' \Biggl[ (1 \!-\! \epsilon)^2 \Bigl\Vert
\vec{x} \!-\! \vec{x}' \Bigr\Vert^2 \!-\! \Bigl( -\frac1{Ha} \!+\!
\frac1{H'a'} \!-\! i \delta\Bigr)^2 \Biggr] \; . \label{dumber}
\end{equation}
Expressions (\ref{dumb}-\ref{dumber}) are well defined for all
values of $\epsilon$ and $D$ except for the two discrete series,
\begin{equation}
\epsilon = \frac{2N}{D \!-\! 2 \!+\! 2N} \qquad {\rm or} \qquad
\epsilon = 1 + \frac{D \!-\! 2}{D \!+\! 2N} \; . \label{discrete}
\end{equation}
Comparison with expression (\ref{smallk}) reveals that these are
just the values for which either the leading small $k$ contribution
from the mode functions --- or one of its $k^{2N}$ corrections
--- combines with the $k^{D-2} dk$ from the measure to give the
$dk/k$ needed to produce a logarithmic infrared divergence
\cite{JMPW}.

This example is perfect for our purposes because expressions
(\ref{dumb}-\ref{dumber}) are how a mathematical physicist would
define the MMC scalar propagator ({\it and the spin two structure
function of the graviton propagator}) for a constant $\epsilon$
cosmology. For $\epsilon > 0$ this example also avoids the charged
issue of de Sitter invariance: there are no symmetries but
homogeneity and isotropy, nor can there be any appeal to the full de
Sitter manifold. And the analogy with expressions (\ref{opensum})
and (\ref{mathsol}-\ref{moremath}) could hardly be clearer: even
though the original mode sum (\ref{naive}) harbors infrared
divergences throughout the range $0 \leq \epsilon \leq 2(D-1)/D$,
these have been automatically subtracted in
(\ref{dumb}-\ref{dumber}), except for the discrete cases
(\ref{discrete}). At those values our mathematically-minded
colleague would say that there is an infrared divergence which
precludes the assumption (\ref{epsmodes}) of minimum energy in the
distant past. But he would insist that (\ref{dumb}) must be the
correct propagator because it solves the propagator equation away
from these values. Before discussing what is wrong with that view we
should follow our own injunction by describing why the infrared
divergence of Ford and Parker happens and how to fix it.

The unphysical thing about (\ref{naive}) is that it assumes each
mode of the initial state was prepared so that it is minimum energy
in the distant past (\ref{epsmodes}). This is possible for initially
sub-horizon modes but causality precludes a local observer from
having much effect on modes which are {\it initially} super-horizon.
Of course the same obstacle exists for a local observer to guarantee
to prepare the initially super-horizon modes in any other state.
However, it is especially problematic to reach the state which was
minimum energy in the distant past because the occupation number
(relative to the instantaneous ground state) tends to grow like
$N(t,k) \sim [H(t_k) a(t)/2k]^2$ \cite{TW5}, so it is quite highly
excited for small $k$, and becoming more so rapidly. The chance of
accidentally hitting this state for a dense set of super-horizon
modes is zero. Hence the naive mode sum (\ref{naive}) represents an
unphysical question which should be reformulated sensibly rather
than being blindly defined by analytic continuation.

Two plausible fixes have been proposed to the infrared problem of
Ford and Parker:
\begin{itemize}
\item{One can retain infinite space in (\ref{naive}) but assume
initial values for the super-horizon modes which are less singular
than (\ref{epsmodes}) \cite{AV}. Of course the time dependence of
the mode functions is determined by the scalar field equation but
their initial values and those of their first time derivatives can
be freely specified. As long as there is no infrared divergence
initially then none can develop \cite{Fulling}.}
\item{One can also work on a compact spatial manifold, such as
the torus $T^{D-1}$, for which there are no initially super-horizon
modes \cite{TW6}. Doing this changes (\ref{naive}) from an integral
to a sum, but it is generally valid to approximate this sum as an
integral with a nonzero lower limit.}
\end{itemize}
In practice there is not much difference between the two fixes in
that they both cut off initially super-horizon modes. Note that both
fixes endow the propagator with a secular dependence associated with
the time elapsed from the initial value surface \cite{JMPW}.

\subsection{Distinguishing Green's functions from
propagators}\label{distinguishing}

The problem with using expression (\ref{dumb}-\ref{dumber}) for the
propagator of a MMC scalar on a constant $\epsilon$ geometry is that
it doesn't represent the expectation value of the time-ordered
product of $\varphi(x) \varphi(x')$ in the presence of any
normalizable state. The same comments apply to using
(\ref{mathsol}-\ref{moremath}) for the propagator of a tachyonic
scalar on de Sitter. Neither statement should come as any surprise.
In both cases direct examination of the mode sums --- (\ref{naive})
and (\ref{opensum}) --- show infrared divergences, and in both cases
analytic regularizations were used to automatically subtract the
power law divergences. {\it These infrared divergences are not
required to solve the propagator equation but they are necessary to
make the result a propagator.} And note that the real problem is
making {\it subtractions}, which corresponds to adding negative (and 
sometimes imaginary) norm states. The subtractions are finite in
closed coordinates, but they are equally invalid.

It should be obvious that there are many solutions to the propagator
equation which are not true propagators. For example, consider $i/2$
times the sum of the advanced and retarded propagators.
Distinguishing a solution to the propagator equation from a true
propagator can sometimes be difficult, especially if one is more
interested in certain symmetries than in physics.

An illuminating example is the one dimensional point particle $q(t)$
of mass $m$ in an inverted potential $V(q) = -\frac12 m \Omega^2
q^2$. This system is especially important because it possesses no
infrared divergence, just like the closed coordinate mode sums. One 
can clearly see how the analytic continuation $\omega \rightarrow 
-i \Omega$ from the conventional harmonic oscillator violates the 
quantum mechanical requirement that states have positive norm. 

The full solution of the Heisenberg operator equation of motion is,
\begin{equation}
q(t) = q_0 \cosh(\Omega t) + \frac{\dot{q}_0}{\Omega} \,
\sinh(\Omega t) \; .
\end{equation}
Of course this system must show quantum mechanical spread. That is
evident from the way its propagator breaks time translation
invariance,
\begin{eqnarray}
\lefteqn{\Bigl\langle \psi \Bigl\vert T\Bigl[ q(t) q(t')\Bigr]
\Bigr\vert \psi \Bigr\rangle = -\frac{i \hbar}{2m \Omega} \,
\sinh\Bigl[ \Omega \vert t \!-\! t'\vert \Bigr] \!+\! \Bigl\langle
\psi \Bigl\vert q_0^2 \Bigr\vert \psi \Bigr\rangle \cosh(\Omega t)
\cosh(\Omega t') } \nonumber \\
& & \hspace{.3cm} + \Bigl\langle \psi \Bigl\vert \frac{q_0 \dot{q}_0
\!+\! \dot{q}_0 q_0}{2 \Omega} \Bigr\vert \psi \Bigr\rangle
\sinh\Bigl[ \Omega (t \!+\! t') \Bigr] \!+\! \Bigl\langle \psi
\Bigl\vert \frac{\dot{q}_0^2}{\Omega^2} \Bigr\vert \psi \Bigr\rangle
\sinh(\Omega t) \sinh(\Omega t') \; . \qquad \label{trueprop}
\end{eqnarray}
Let us denote the three expectation values by the letters $A$, $B$
and $C$,
\begin{equation}
A \equiv \Bigl\langle \psi \Bigl\vert q_0^2 \Bigr\vert \psi
\Bigr\rangle \quad , \quad B \equiv \Bigl\langle \psi \Bigl\vert
\frac{q_0 \dot{q}_0 \!+\! \dot{q}_0 q_0}{2 \Omega} \Bigr\vert \psi
\Bigr\rangle \quad , \quad C \equiv \Bigl\langle \psi \Bigl\vert
\frac{\dot{q}_0^2}{\Omega^2} \Bigr\vert \psi \Bigr\rangle \; .
\end{equation}
For large $t$ and $t'$ we can express the real part of
(\ref{trueprop}) as,
\begin{eqnarray}
\lefteqn{A \cosh(\Omega t) \cosh(\Omega t') \!+\! B \sinh\Bigl[
\Omega (t \!+\! t') \Bigr] \!+\! C \sinh(\Omega t) \sinh(\Omega t')
} \nonumber \\
& & \hspace{7cm} \longrightarrow \frac14 e^{\Omega (t + t')} \Bigl\{
A \!+\! 2 B \!+\! C\Bigr\} \; . \qquad \label{ABC}
\end{eqnarray}
It is simple to see that the constant $A + 2 B + C$ must be
positive. First, note $A > 0$ and $C > 0$ because they are the
expectation values of positive operators. Only the constant $B$
might be negative. Hence we can write,
\begin{equation}
A + 2 B + C \geq A - 2 \sqrt{B^2} + C \; . \label{step1}
\end{equation}
Now note that $A$, $B$ and $C$ are constrained by the Uncertainty
Principle and by the Schwarz Inequality (with the requirement of
normalizability),
\begin{equation}
A \times C \geq \frac{\hbar^2}{4 m^2 \Omega^2} \qquad , \qquad A
\times C > B^2 \; . \label{QMIDS}
\end{equation}
Using the second relation of (\ref{QMIDS}) in (\ref{step1}) allows
us to reach the desired conclusion after some simple algebra,
\begin{equation}
A + 2 B + C > A - 2 \sqrt{A C} + C = \Bigl( \sqrt{A} - \sqrt{C}
\Bigr)^2 \geq 0 \; . \label{step2}
\end{equation}
Hence the propagator (\ref{trueprop}) must show secular growth which
violates time translation invariance for any valid quantum
mechanical state $\vert \psi \rangle$.

The preceding paragraph is how a quantum physicist would go about
solving the propagator equation,
\begin{equation}
-m \Bigl[\frac{d^2}{dt^2} \!-\! \Omega^2 \Bigr] i\Delta(t;t') =
i\hbar \delta(t \!-\! t') \; . \label{1Deqn}
\end{equation}
However, a mathematical physicist might consider (\ref{1Deqn}) to be
simply a second order differential equation he can solve at will. He
might be very attracted to the solution that follows from making the
analytic continuation $\omega \rightarrow -i \Omega$ in the
propagator of the simple harmonic oscillator,
\begin{equation}
\frac{\hbar}{2 m \omega} \, e^{-i \omega \vert t - t'\vert} \qquad
\longrightarrow \qquad \frac{i \hbar}{2 m \Omega} \, e^{-\Omega
\vert t - t'\vert } \; . \label{idiotic}
\end{equation}
The right hand side of expression (\ref{idiotic}) solves the
propagator equation (\ref{1Deqn}); it is also time translation
invariant, and it falls off exponentially as the difference between
$t$ and $t'$ grows. However, comparison with expression
(\ref{trueprop}) reveals some very peculiar quantum mechanics,
\begin{equation}
\frac{i \hbar}{2 m \Omega} \, e^{-\Omega \vert t - t'\vert } \qquad
\Longrightarrow \qquad A = -C = \frac{i \hbar}{2 m \Omega} \quad ,
\quad B = 0 \; . \label{impossible}
\end{equation}
There is no normalizable state $\vert \psi \rangle$ for which
positive operators such as $q_0^2$ and $\dot{q}_0^2$ can have
imaginary expectation values. Expression(\ref{idiotic}) is a formal
solution to the propagator equation which isn't a true propagator,
even though it derives from analytic continuation ($\omega
\rightarrow -i\Omega$) of a true propagator. The same comments
pertain to expressions (\ref{dumb}-\ref{dumber}), for $0 \leq
\epsilon \leq 2 (D-1)/D$, and to expressions
(\ref{mathsol}-\ref{moremath}), for $M^2 \leq 0$.

\subsection{Math versus physics}

We have seen that the application of analytic continuation to an
infrared singular mode sum --- such as (\ref{opensum}) or
(\ref{naive}) --- whose order of divergence depends upon some free
parameter --- such as the mass-squared \cite{MTW2}, the dimension of
spacetime \cite{MTW2,KK}, or the cosmological deceleration parameter
\cite{JMPW} --- will only reveal divergences for the discrete,
special values of the parameter that happen to produce logarithmic
divergences. These special values always abut, at least on one side,
a continuum for which analytic continuation is equally invalid on
account of power law divergences that simply fail to show up in the
analytic continuation. In this case the use of analytic continuation
produces a formal solution to the propagator equation which is not a
propagator in the sense of being the expectation value of the
time-ordered product of two fields in the presence of a normalizable
state \cite{TW3}.

This has been pointed out before \cite{MTW2} but the conclusion is
not uniformly accepted \cite{Higuchi}. Indeed, Morrison claims to
have verified the results of analytic continuation in the signature
by demonstrating that they agree with analytic continuation in the
mass-squared \cite{IAM}. Of course what he actually showed is that
both analytic continuations make the same error of automatically
subtracting power law infrared divergences. Some mathematical 
physicists also attempt to justify their analytic continuations in
closed coordinates, where there are no infrared divergences. We will
treat that in detail in the next section.

To close this section it is interesting, and in a sense more 
powerful, to briefly discuss the phenomenological consequences that 
would follow if the mathematical viewpoint was to be accepted. The 
basic problem is that tachyonic scalars roll down their potentials, 
even in de Sitter space. While this is admitted for $M^2 = -N
(N + D-1) H^2$, it is denied for all other tachyonic masses. The 
conclusion would be that we have a physical theory with the bizarre 
feature that a scalar with one of these special masses does decay, 
but (it is claimed that) the decay can be stabilized by making the 
mass a little {\it more} tachyonic!

Much worse follows when we turn to the Standard Model Higgs scalar
whose tachyonic mass term is responsible for spontaneous symmetry
breaking. Consider the Gedanken experiment of formulating the
Standard Model in the symmetric vacuum on de Sitter background. Does
the Higgs field roll down its tachyonic potential to break $SU(2)
\times U(1)$ and give mass to the $W^{\pm}$ and $Z^0$ bosons, along
with the quarks and charged leptons? Note that we can make the de
Sitter Hubble constant enormously smaller than the magnitude of the
tachyonic Higgs mass. For example, in the current universe it would
be 44 orders of magnitude smaller. In this context the claim implies
that the Higgs would not roll down its potential unless the
tachyonic mass happens to agree with one of the discrete values $M^2
= -N (N + D-1) H^2$! So spontaneous symmetry breaking would be
controlled by the gravitational parameter $H$ whose scale is 44
orders of magnitude below the electro-weak scale! Furthermore,
minuscule fractional changes in the Higgs mass would lead to
completely different physics.

\section{Zero Modes Are Not the Problem}\label{real}

Mathematical physicists distrust open coordinates because they do 
not cover the full de Sitter manifold. They suspect that the naive 
use of open coordinates has misled cosmologists into making subtle 
errors, and that a clearer picture emerges in closed coordinates. In 
particular, the closed coordinate mode functions are discrete so 
there can be nothing like the accumulation of very small Fourier $k$ 
modes which leads to the infrared divergence of the open coordinate 
mode sums for the MMC scalar and graviton propagators. Moreover, the 
belief is that the infrared divergence of the MMC scalar propagator 
is reflected, in closed coordinates, by the fact that there is a 
discrete zero mode. Because the graviton has no such zero mode, they 
argue that there can be no problem with the graviton propagator.

This view of the genesis of de Sitter breaking in the MMC scalar has
long been recognized as false. Let us quote from the classic 1985
discussion by Bruce Allen \cite{BA}:
\begin{quotation}
{\it It is often believed that ``what goes wrong'' when $m^2 = 0$
has something to do with the fact that the wave equation has a
constant solution, which is often called a ``zero mode.'' This is
simply not true.}
\end{quotation}
In fact the problem with the massless, minimally coupled scalar is
not its single zero mode but rather the way {\it all} modes behave
at late times. This emerges clearly from equation (4.5) of the paper
by Allen and Folacci \cite{AF}, in which the zero mode is excluded
from the scalar mode sum in $D=4$ closed coordinates. In our
notation this relation reads,
\begin{eqnarray}
\lefteqn{G^{(1)}_{\rm NZM}(x;x') } \nonumber \\
& & \equiv 2 {\rm Re}\Biggl[ \sum_{k =1}^{\infty} \sum_{\ell = 0}^k
\sum_{m = -\ell}^{\ell} u_{k}(\tau) {\rm Y}_{k\ell
m}(\chi,\theta,\phi) \!\times\! u^*_{k}(\tau') {\rm Y}^*_{k\ell
m}(\chi',\theta',\phi') \Biggr] \; ,
\qquad \\
& & = \frac{H^2}{4 \pi^2} \Biggl[ \frac{4}{y} \!-\! \ln\Bigl[
\frac{y(x;x')}{\cosh(H \tau) \cosh(H \tau')} \Bigr] \!-\! {\rm
sech}^2(H \tau) \!-\! {\rm sech}^2(H \tau') \Biggr] \; . \label{LOG1}
\end{eqnarray}

The de Sitter breaking logarithm in the coincidence limit does not
arise from the zero mode but rather from the late time limiting form
which {\it all} modes approach,
\begin{equation}
u_{k}(\tau) \longrightarrow \frac{H}{2\pi k^{\frac32}} \; .
\label{limit}
\end{equation}
This form sets in for $H \tau \gtwid \ln(k)$, before which there is
destructive interference from oscillations, so one is effectively
summing $1/k$ up to $k \sim e^{H\tau}$,
\begin{equation}
\sum_{k =1}^{e^{H \tau}} \sum_{\ell = 0}^k \sum_{m = -\ell}^{\ell} 
\Bigl\vert u_{k}(\tau) {\rm Y}_{k\ell m}(\chi,\theta,\phi) \Bigr\vert^2
\approx \int_1^{e^{H \tau}} \!\!\!\! dk \, k^2 \times 
\frac{H^2}{4 \pi^2 k^3} = \frac{H^2}{4 \pi^2} \, H\tau \; . \label{LOG2}
\end{equation}
Because $u_k(\tau)$ and $u^*_k(\tau')$ both approach the same form 
(\ref{limit}), each mode contributes {\it positively}. One can only 
avoid the de Sitter breaking growth by including negative norm states.

It will be seen that the de Sitter breaking, secular growth of 
expressions (\ref{LOG1}) and (\ref{LOG2}) derives from the fact that 
more and more modes approach the saturated condition (\ref{limit}) as
time progresses. Hence de Sitter breaking originates in the large (but 
still finite) $k$ end of the mode sum, where there is not even any 
distinction between open and closed coordinates. In particular, the 
presence or absence of a zero mode is irrelevant.

Cosmologists ascribe two properties to modes which obey (\ref{limit}): 
\begin{itemize}
\item{{\it Freezing in}; and}
\item{{\it Scale invariance}.}
\end{itemize}
These two properties are why the primordial scalar power spectrum can be
observed during the current epoch, so no amount of clever mathematics can
make them disappear. Graviton mode functions possess the same key 
features of freezing in and scale invariance. Hence the closed coordinate 
mode sum for the coincidence limit of the graviton propagator must 
possess the same de Sitter breaking infrared logarithm. This is not some 
sort of gauge artifact, it is precisely why the tensor power spectrum 
from primordial inflation can be observed during the current epoch. In
section \ref{spin2} we will see that the presence of this de Sitter
breaking infrared logarithm is not being recognized by mathematical
physicists partly because they substitute formal solutions for the
original, de Sitter breaking mode sums and partly because they
employ analytic continuation techniques to evaluate those mode sums
they do consider.

Before concluding we should comment on the closed coordinate mode sum 
(\ref{closedsum}) for tachyons. This is free of infrared divergences, 
which has prompted some mathematical physicists to claim that it fails 
to show de Sitter breaking \cite{Higuchi}. That is incorrect. The de 
Sitter breaking of the closed coordinate mode sum derives from more and 
more modes approaching the saturated condition (\ref{lateclosed}) as 
time progresses. Because each mode contributes positively, there is no
way to avoid this without violating the canonical normalization 
condition (\ref{closedWron}) that all states have positive norm. 
Analytically continuing from $M^2$ positive to negative represents
such a violation, just as we saw with the equally illegitimate analytic 
continuation $\omega \rightarrow -i \Omega$ for the inverted harmonic 
potential (\ref{impossible}). In fact, the use of negative norm states 
to produce a de Sitter invariant solution to the tachyon propagator 
equation is admitted by Faisal and Higuchi \cite{Higuchi}:
\begin{quotation}
{\it We note in passing that the modes $\Phi^{(\ell \ell_2 m)}$ with 
positive norm form a unitary representation of the de Sitter group if 
$L_0$ is an integer whereas for a positive non-integer value of $L_0$ 
no unitary representation exists because of the negative norm modes} 
\end{quotation}
The authors do not seem to have realized that this admission precludes
their solution from being a true propagator.

Faisal and Higuchi are also wrong in employing the term ``infrared
divergence'' to describe what happens for $M^2 = -N (N + D -1) H^2$, 
which is the case of their quantity $L_0$ being a nonnegative 
integer. The problem actually arises from their analytically 
continued mode functions becoming degenerate. These mode functions
consist of powers of the scale factor times associated Legendre 
polynomials $P^{\mu}_{\nu}(z)$ evaluated at $z = i \sinh(H \tau)$ 
\cite{MM,AH}. Because the associated Legendre polynomial is evaluated 
at an imaginary argument, the mode function and its complex conjugate 
are linearly independent for most values of $M^2$, leading to a 
nonzero (although negative) Wronskian. When $M^2 = -N (N + D - 1) H^2$ 
the mode function becomes proportional to its complex conjugate so that 
the Wronskian between them vanishes the same way it does for 
$J_{\nu}(z)$ and $J_{-\nu}(z)$ when $\nu$ becomes an integer. That 
could have been avoided by employing the second linearly independent 
solution, $Q^{\mu}_{\nu}(z)$, along with $P^{\mu}_{\nu}(z)$, whose 
peculiar time dependence would make de Sitter breaking even more 
obvious.

\section{Spin 2 Sector of the Graviton
Propagator}\label{spin2}

One of the nice features of Morrison's paper is that he has
identified precisely where the two approaches diverge in
constructing the graviton propagator when a de Sitter invariant
gauge condition is imposed as a strong operator
equation,\footnote{For $\beta = 2$ condition (\ref{gauge}) cannot be
imposed because it implies the vanishing of the linearized Ricci
scalar, which is gauge invariant at linearized order.}
\begin{equation}
g^{\rho\sigma} \Bigl[ h_{\mu\rho ; \sigma} - \frac{\beta}2
h_{\rho\sigma ; \mu} \Bigr] = 0 \; . \label{gauge}
\end{equation}
In that case the propagator consists of a spin zero part which
derives from the constrained part of the gravitational field and a
transverse-traceless (spin two) sector which derives from the
$\frac12 D (D-3)$ dynamical gravitons and the remaining $(D-1)$
constrained fields. The spin zero structure function involves the
scalar propagator for $M^2 = -2(D-1) H^2/(2 -\beta)$, which is
infrared singular and de Sitter breaking for all $\beta < 2$
\cite{PMTW1}. The comments of section~\ref{tachyon} have already
addressed the curious contention that there is no de Sitter breaking
except for the discrete values of $\beta = 2 -2(D-1)/N(N+D-1)$
\cite{Higuchi,IAM}. In this section we discuss what Morrison's work
says about the difference between the two approaches regarding the
spin two sector.

\subsection{The price of de Sitter invariance}\label{price}

This subsection begins by summarizing notation. Then we review the
derivation employed \cite{MTW3} for the de Sitter breaking solution
to the spin two sector of the propagator. The subsection closes by
identifying the two points at which Morrison's de Sitter invariant
construction deviates from ours.

In any coordinate system we define the graviton field $h_{\mu\nu}$
by subtracting the de Sitter background $g_{\mu\nu}$ from the full
metric,
\begin{equation}
g_{\mu\nu}^{\rm full} \equiv g_{\mu\nu} + \kappa h_{\mu\nu} \qquad ,
\qquad \kappa^2 \equiv 16 \pi G \; . \label{hdef}
\end{equation}
By convention its indices are raised and lowered with the de Sitter
background metric. Covariant derivatives with respect to the de
Sitter background are represented by $D_{\alpha}$ and $\square
\equiv g^{\alpha\beta} D_{\alpha} D_{\beta}$.

The spin two part of the graviton propagator takes the form
\cite{MTW3},
\begin{equation}
i\Bigl[\mbox{}_{\mu\nu} \Delta^2_{\rho\sigma}\Bigr](x;x') = \frac1{4
H^4} \mathbf{P}_{\mu\nu}^{~~\alpha\beta}(x) \times
\mathbf{P}_{\rho\sigma}^{~~\kappa\lambda}(x') \Bigl[
\mathcal{R}_{\alpha\kappa}(x;x') \mathcal{R}_{\beta\lambda}(x;x')
\mathcal{S}_2(x;x')\Bigr] \; , \label{Spin2}
\end{equation}
where $\mathcal{S}_2(x;x')$ is the spin two structure function,
$\mathcal{R}_{\alpha\kappa}(x;x')$ is a mixed second derivative of
the de Sitter length function $y(x;x')$ (\ref{moremath}), normalized
to give $g_{\alpha\kappa}$ in the coincidence limit,
\begin{equation}
\mathcal{R}_{\alpha\kappa}(x;x') \equiv -\frac1{2 H^2}
\frac{\partial^2 y(x;x')}{\partial x^{\alpha} \partial
{x'}^{\kappa}} \qquad \Longrightarrow \qquad
\mathcal{R}_{\alpha\kappa}(x;x) = g_{\alpha\kappa}(x) \; ,
\end{equation}
and ${\bf P}_{\mu\nu}^{~~ \alpha\beta}(x)$ is the
transverse-traceless projector,
\begin{eqnarray}
\lefteqn{\mathbf{P}_{\mu\nu}^{~~\alpha\beta} \equiv \frac12
\Bigl(\frac{D\!-\!3}{D\!-\!2}\Bigr) \Biggl\{ -\delta^{\alpha}_{(\mu}
\delta^{\beta}_{\nu)} \Bigl[\square \!-\! D H^2\Bigr] \Bigl[ \square
\!-\! 2 H^2\Bigr] + 2 D_{(\mu} \Bigl[ \square \!+\! H^2\Bigr]
\delta^{(\alpha}_{\nu)} D^{\beta)} } \nonumber \\
& & \hspace{.3cm} - \Bigl( \frac{D \!-\!2}{D \!-\!1} \Bigr) D_{(\mu}
D_{\nu)} D^{(\alpha} D^{\beta)} + g_{\mu\nu} g^{\alpha\beta} \Bigl[
\frac{\square^2}{D \!-\! 1} \!-\! H^2 \square \!+\! 2 H^4\Bigr]
\qquad \nonumber \\
& & \hspace{.3cm} -\frac{D_{(\mu} D_{\nu)} }{D \!-\! 1} \Bigl[
\square \!+\! 2 (D \!-\! 1) H^2\Bigr] g^{\alpha\beta}
-\frac{g_{\mu\nu} }{D \!-\! 1} \Bigl[ \square \!+\! 2 (D \!-\! 1)
H^2\Bigr] D^{(\alpha} D^{\beta)} \Biggr\} . \qquad \label{spin2op}
\end{eqnarray}
Our form (\ref{Spin2}) is preferable to the representation employed
in the mathematical physics literature \cite{INVPROP} because its
tensor structure makes no assumption of de Sitter invariance and
because the essential spacetime dependence is represented using only
a {\it single} scalar structure function $\mathcal{S}_2(x;x')$,
rather than having a distinct scalar coefficient function for each
of the five de Sitter invariant tensor factors.\footnote{If one
imposes only the cosmological symmetries of homogeneity and isotropy
the number of tensor factors rises to 14 \cite{KMW}.} It is also
worth pointing out that this representation could be generalized to
any background if we note that the transverse-traceless projector is
${\bf P}_{\mu\nu}^{~~ \alpha\gamma} \equiv \mathcal{P}_{\mu\nu}^{~~
\alpha\beta\gamma\delta} D_{\beta} D_{\delta}$ \cite{MTW3}, where
the second order differential operator $\mathcal{P}_{\mu\nu}^{~~
\alpha\beta\gamma\delta}$ can be read off from the linearized Weyl
tensor \cite{SPW},
\begin{equation}
C^{\alpha\beta\gamma\delta} = \mathcal{P}_{\mu\nu}^{~~
\alpha\beta\gamma\delta} \times h^{\mu\nu} + O(h^2) \; .
\label{Weyl}
\end{equation}

The operator ${\bf P}_{\mu\nu}^{~~ \alpha\beta}$ has four important
properties. The first two are transversality and tracelessness on
each of its index groups \cite{MTW3},
\begin{equation}
g^{\mu\nu} \times {\bf P}_{\mu\nu}^{~~ \alpha\beta} = 0 = {\bf
P}_{\mu\nu}^{~~ \alpha\beta} \times g_{\alpha\beta} \quad , \quad
D^{\mu} \times {\bf P}_{\mu\nu}^{~~ \alpha\beta} = 0 = {\bf
P}_{\mu\nu}^{~~ \alpha\beta} \times D_{\alpha} \; . \label{ID1}
\end{equation}
The third property is commuting with the d'Alembertian \cite{MTW3},
\begin{equation}
\square \times {\bf P}_{\mu\nu}^{~~ \alpha\beta} = {\bf
P}_{\mu\nu}^{~~ \alpha\beta} \times \square \; . \label{ID3}
\end{equation}
And the final property concerns its square \cite{MTW3},
\begin{equation}
{\bf P}_{\mu\nu}^{~~ \gamma\delta} \times {\bf P}_{\gamma\delta}^{~~
\alpha\beta} = -\frac12 \Bigl( \frac{D \!-\! 3}{D \!-\! 2}\Bigr)
\Bigl[\square \!-\! 2 H^2\Bigr] \Bigl[ \square \!-\! D H^2 \Bigr]
{\bf P}_{\mu\nu}^{~~ \alpha\beta} \; . \label{ID4}
\end{equation}
The product of the transverse projector and two factors of
$\mathcal{R}$ also obeys an important commutation relation
\cite{MTW3},
\begin{equation}
\square {\bf P}_{\mu\nu}^{~~ \alpha\beta}(x)
\mathcal{R}_{\alpha\kappa}(x;x') \mathcal{R}_{\beta\lambda}(x;x') =
{\bf P}_{\mu\nu}^{~~ \alpha\beta}(x)
\mathcal{R}_{\alpha\kappa}(x;x') \mathcal{R}_{\beta\lambda}(x;x')
\Bigl[ \square \!+\! 2 H^2\Bigr] \; . \label{ID5}
\end{equation}

The spin two part of the propagator equation reads \cite{MTW3},
\begin{equation}
\frac12 \Bigl[ \square \!-\! 2 H^2\Bigr] \, i\Bigl[\mbox{}_{\mu\nu}
\Delta^2_{\rho\sigma}\Bigr](x;x') = i \Bigl[\mbox{}_{\mu\nu}
P^2_{\rho\sigma} \Bigr](x;x') \; . \label{spin2eqn}
\end{equation}
The quantity on the right hand side of (\ref{spin2eqn}) is $i$ times
the transverse-traceless projection operator. It takes the same form
(\ref{Spin2}) as the spin two part of the graviton propagator but
with a different structure function $\mathcal{P}_2(x;x')$,
\begin{equation}
i\Bigl[\mbox{}_{\mu\nu} P^2_{\rho\sigma}\Bigr](x;x') = \frac1{4 H^4}
\mathbf{P}_{\mu\nu}^{~~\alpha\beta}(x) \times
\mathbf{P}_{\rho\sigma}^{~~\kappa\lambda}(x') \Bigl[
\mathcal{R}_{\alpha\kappa}(x;x') \mathcal{R}_{\beta\lambda}(x;x')
\mathcal{P}_2(x;x')\Bigr] \; . \label{Proj2}
\end{equation}
It can also be expressed as,
\begin{equation}
i\Bigl[\mbox{}_{\mu\nu} P^2_{\rho\sigma}\Bigr](x;x') = g_{\mu (\rho}
g_{\sigma) \nu} \times \frac{ i \delta^D(x \!-\! x')}{\sqrt{-g}} +
\Bigl( {\rm Traces\ and\ Gradients} \Bigr) \; . \label{altP2}
\end{equation}
where the traces and gradients enforce transversality and traceless.

Acting transverse-traceless projectors on the $x^{\mu}$ and
${x'}^{\mu}$ dependence of expression (\ref{altP2}) and exploiting
relations (\ref{ID1}-\ref{ID5}) gives an equation for the structure
function $\mathcal{P}_2(x;x')$,
\begin{eqnarray}
\lefteqn{{\bf P}_{\mu\nu}^{~~ \alpha\beta}(x) \!\times\! {\bf
P}_{\rho\sigma}^{~~ \kappa\lambda}(x') \Biggl\{
\mathcal{R}_{\alpha\kappa} \mathcal{R}_{\beta\lambda} \square \Bigl[
\square \!-\! (D\!-\! 2) H^2\Bigr] \square' \Bigl[ \square' \!-\! (D
\!-\! 2) H^2\Bigr] \mathcal{P}_2 \Biggr\} } \nonumber \\
& & \hspace{1cm} = {\bf P}_{\mu\nu}^{~~ \alpha\beta}(x) \!\times\!
{\bf P}_{\rho\sigma}^{~~ \kappa\lambda}(x') \Biggl\{
\mathcal{R}_{\alpha \kappa} \mathcal{R}_{\beta\lambda} \times 16
\Bigl(\frac{D \!-\! 2}{D \!-\! 3}\Bigr)^2 H^4 \frac{i \delta^D(x
\!-\! x')}{\sqrt{-g}} \Biggr\} . \qquad \label{almost}
\end{eqnarray}
If it is valid to drop the projectors from both sides of
(\ref{almost}) we would have a scalar equation for
$\mathcal{P}_2(x;x')$ \cite{MTW3},
\begin{equation}
\square \Bigl[ \square \!-\! (D\!-\! 2) H^2\Bigr] \square' \Bigl[
\square' \!-\! (D \!-\! 2) H^2\Bigr] \mathcal{P}_2(x;x') = 16
\Bigl(\frac{D \!-\! 2}{D \!-\! 3}\Bigr)^2 H^4 \frac{i \delta^D(x
\!-\! x')}{\sqrt{-g}} \; . \label{disputed}
\end{equation}
Equation (\ref{disputed}) implies that $[\square - (D-2) H^2]
\square' [\square' - (D-2) H^2] \mathcal{P}_2(x;x')$ is proportional
to the de Sitter breaking propagator of the MMC scalar, so
$\mathcal{P}_2(x;x')$ would necessarily break de Sitter invariance
as well.

It is straightforward to derive explicit solutions to equations such
as (\ref{disputed}). The trick is to consider the scalar propagator
$\Delta_i(x;x')$ for an arbitrary mass-squared $M_i^2$,
\begin{equation}
\Bigl[ \square \!-\! M_i^2 \Bigr] i\Delta_i(x;x') = \frac{i
\delta^D(x \!-\! x')}{\sqrt{-g}} \; .
\end{equation}
Then if acting $[\square - M_1^2]$ on $i\Delta_{12}(x;x')$ produces
the propagator $i\Delta_2(x;x')$, the solution is straightforward
\cite{MTW2},
\begin{equation}
\Bigl[ \square \!-\! M_1^2 \Bigr] i\Delta_{12}(x;x') =
i\Delta_2(x;x') \qquad \Longrightarrow \qquad i\Delta_{12} =
\frac{i\Delta_1 \!-\! i\Delta_2}{M_1^2 \!-\! M_2^2} \; .
\label{inttrick}
\end{equation}
The same trick works when the source is an integrated propagator,
\begin{equation}
\Bigl[ \square \!-\! M_1^2 \Bigr] i\Delta_{123}(x;x') =
i\Delta_{23}(x;x') \qquad \Longrightarrow \qquad i\Delta_{123} =
\frac{i\Delta_{12} \!-\! i\Delta_{13}}{M_2^2 \!-\! M_3^2} \; .
\label{inttrick2}
\end{equation}
When two of the masses coincide one gets a derivative with respect
to mass-squared. {\it Note, however, that these relations require
one to consider the scalar propagator $i\Delta_i(x;x')$ as an
analytic function of its mass-squared $M_i^2$.} As we have explained
in section~\ref{tachyon}, that assumption of analyticity in $M_i^2$
is only valid if one allows the propagators to break de Sitter
invariance when $M_i^2$ goes from positive to negative, so de Sitter
breaking must be evident even for $M^2_i$ slightly positive.

One can show that the de Sitter breaking of $\mathcal{P}_2(x;x')$
which is implied by equation (\ref{disputed}) does not drop out of
the spin two projection operator (\ref{Proj2}) \cite{KMW}. At this
point it is obvious from equation (\ref{spin2eqn}) that the spin two
sector of the graviton propagator must break de Sitter invariance as
well. Indeed, the same manipulations that led to (\ref{disputed})
give,
\begin{equation}
\frac12 \square \mathcal{S}_2(x;x') = \mathcal{P}_2(x;x') \; .
\label{disp2}
\end{equation}
and acting $\square [\square - (D-2) H^2] \square' [\square' - (D-2)
H^2]$ on both sides gives,
\begin{equation}
\square^2 \Bigl[ \square \!-\! (D\!-\! 2) H^2\Bigr] \square' \Bigl[
\square' \!-\! (D \!-\! 2) H^2\Bigr] \mathcal{S}_2(x;x') = 32
\Bigl(\frac{D \!-\! 2}{D \!-\! 3}\Bigr)^2 H^4 \frac{i \delta^D(x
\!-\! x')}{\sqrt{-g}} \; . \label{disp3}
\end{equation}
The de Sitter breaking implied for $i[\mbox{}_{\mu\nu}
\Delta^2_{\rho\sigma}](x;x')$ by (\ref{disputed}-\ref{disp3}) has
been explicitly worked out \cite{KMW} and shown to agree with both
the noncovariant gauge propagator \cite{TW1} and with the result in
transverse-traceless-spatial gauge \cite{MTW4}.

This completes our review of how the de Sitter breaking solution was
constructed \cite{MTW3}. Morrison has demonstrated that the de
Sitter invariant solutions \cite{INVPROP,Higuchi} follow by
deviating from this procedure at two points \cite{IAM}:
\begin{enumerate}
\item{One must add a special constant to the right hand side of equation
(\ref{disp3}) --- or equivalently, to the right hand side of
equation (\ref{disputed}); and}
\item{One must solve integrated propagator equations of the form
(\ref{inttrick}-\ref{inttrick2}) by assuming that the de Sitter
invariant scalar propagator is a meromorphic function of its
mass-squared, with simple poles at $M^2 = -N (N + D -1) H^2$.}
\end{enumerate}
We have already explained in section~\ref{tachyon} that the second
deviation produces formal solutions to the desired equations which
are not true propagators. That single observation would suffice to
invalidate the mathematical physics solutions, but it happens that
the first deviation is also problematic.

\subsection{Why no constant can be added}\label{whyno}

The motivation for the first deviation is the fact, noted in earlier
work \cite{KMW}, that the transverse-traceless projectors annihilate
constant shifts in the structure functions,
\begin{equation}
{\bf P}_{\mu\nu}^{~~ \alpha\beta}(x) \times {\bf P}_{\rho\sigma}^{~~
\kappa\lambda}(x') \Bigl[ \mathcal{R}_{\alpha\kappa}(x;x')
\mathcal{R}_{\beta\lambda}(x;x') \Bigr] = 0 \; . \label{constant}
\end{equation}
Hence, it is claimed, one cannot pass from equation (\ref{almost})
to (\ref{disputed}) \cite{IAM}. If we were only interested in
solving differential equation (\ref{spin2eqn}) this conclusion would
be correct. However, what we really seek is a propagator, and
section~\ref{tachyon} has already demonstrated that propagator
equations have many solutions that are not true propagators
\cite{TW3}. When constructing a propagator one can indeed pass from
equation (\ref{almost}) to (\ref{disputed}).

The simplest way to see what is wrong with adding a constant to
equation (\ref{disputed}) is by taking the flat space limit. In that
limit the two structure functions become translation invariant,
\begin{equation}
\lim_{H \rightarrow 0} \frac{\mathcal{S}_2(x;x')}{4 H^4} \equiv
S_{\rm flt}(x \!-\! x') \qquad , \qquad \lim_{H \rightarrow 0}
\frac{\mathcal{P}_2(x;x')}{4 H^4} \equiv P_{\rm flt}(x \!-\! x') \;
.
\end{equation}
The spin two part of the graviton propagator (\ref{Spin2}) also
takes the simple form,
\begin{eqnarray}
i\Bigl[ \mbox{}_{\mu\nu} \Delta^{\rm flt}_{\rho\sigma}\Bigr](x;x') &
\equiv & \lim_{H \rightarrow 0} i \Bigl[\mbox{}_{\mu\nu}
\Delta^2_{\rho\sigma}\Bigr](x;x') \; , \\
& = & \frac14 \Bigl(\frac{D \!-\! 3}{D \!-\! 2}\Bigr)^2 \Biggl[
\Pi_{\mu (\rho} \Pi_{\sigma) \nu} - \frac{\Pi_{\mu\nu}
\Pi_{\rho\sigma}}{D \!-\! 1} \Biggr] \partial^4 S_{\rm flt}(x \!-\!
x') \; , \qquad \label{flatprop}
\end{eqnarray}
where indices are raised and lowered with the Minkowski metric
$\eta_{\mu\nu}$, parenthesized indices are symmetrized, $\partial^2
\equiv \eta^{\mu\nu} \partial_{\mu} \partial_{\nu}$ is the flat
space d'Alem\-ber\-tian and $\Pi_{\mu\nu}$ is the transverse
projector,
\begin{equation}
\Pi_{\mu\nu} \equiv \eta_{\mu\nu} \partial^2 - \partial_{\mu}
\partial_{\nu} \; .
\end{equation}
The 8th order differential operator acting upon $S_{\rm flt}(x -
x')$ in expression (\ref{flatprop}) appears so frequently in this
discussion that we will denote it by the symbol ${\bf
T}_{\mu\nu\rho\sigma}$,
\begin{equation}
{\bf T}_{\mu\nu\rho\sigma} \equiv \frac14 \Bigl(\frac{D \!-\! 3}{D
\!-\! 2}\Bigr)^2 \Biggl[ \Pi_{\mu (\rho} \Pi_{\sigma) \nu} -
\frac{\Pi_{\mu\nu} \Pi_{\rho\sigma}}{D \!-\! 1} \Biggr] \partial^4
\; . \label{flatproj}
\end{equation}
Of course the factor of $(D - 3)^2$ derives from the two Weyl
tensors (\ref{Weyl}) involved in the construction of ${\bf
T}_{\mu\nu\rho\sigma}$.

The flat space limits of the graviton propagator equation
(\ref{spin2eqn}) and the defining relation (\ref{almost}) for the
transverse-traceless projection operator are,
\begin{eqnarray}
{\bf T}_{\mu\nu\rho\sigma} \times \frac{\partial^2}{2} S_{\rm flt} &
= & {\bf T}_{\mu\nu\rho\sigma} \times P_{\rm flt} \; , \qquad
\label{flateqn1} \\
{\bf T}_{\mu\nu\rho\sigma} \times \partial^8 P_{\rm flt} & = & {\bf
T}_{\mu\nu\rho\sigma} \times 4 \Bigl( \frac{D \!-\! 2}{D \!-\!
3}\Bigr)^2 i \delta^D(x \!-\! x') \; . \qquad \label{flateqn2}
\end{eqnarray}
The point the mathematical physicists dispute is the validity of
removing the factors of ${\bf T}_{\mu\nu\rho\sigma}$ from equations
(\ref{flateqn1}-\ref{flateqn2}) to conclude,
\begin{equation}
\partial^{10} S_{\rm flt}(x \!-\! x') = 8 \Bigl( \frac{D \!-\! 2}{D
\!-\! 3}\Bigr)^2 i\delta^D(x \!-\! x') \; . \label{flatdisp}
\end{equation}
If equation (\ref{flatdisp}) is accepted, the spin two structure
function obeys,
\begin{equation}
\partial^4 S_{\rm flt}(x \!-\! x') = 8 \Bigl( \frac{D \!-\! 2}{D
\!-\! 3}\Bigr)^2 \frac{\Gamma(\frac{D}2 \!-\! 1)}{4 \pi^{\frac{D}2}}
\frac{(\Delta x^{6 - D} \!-\! \mu^{D - 4} \Delta x^2)}{8 (D \!-\! 6)
(D \!-\! 4)} \; , \label{flatsol}
\end{equation}
where $\Delta x^2 \equiv \eta_{\mu\nu} (x - x')^{\mu} (x -
x')^{\nu}$. Substituting this form for the structure function into
(\ref{flatprop}) gives the recognized spin two part of the graviton
propagator in flat space \cite{Capper},
\begin{eqnarray}
\lefteqn{ i\Bigl[ \mbox{}_{\mu\nu} \Delta^{\rm
flt}_{\rho\sigma}\Bigr](x;x') = \frac14 \Bigl( \frac{D \!-\! 2}{D
\!-\! 1} \Bigr) \Biggl\{ 3 \eta_{(\mu\nu} \eta_{\rho\sigma)} - (D
\!+\! 2) \frac{[\eta_{\mu\nu} \Delta_{\rho} \Delta x_{\sigma} \!+\!
\Delta x_{\mu} \Delta x_{\nu} \eta_{\rho\sigma}] }{\Delta x^2} }
\nonumber \\
& & \hspace{.7cm} + 4 D \frac{ \Delta x_{(\mu} \eta_{\nu) (\rho}
\Delta x_{\sigma)} }{\Delta x^2} + D (D \!-\! 2) \frac{\Delta
x_{\mu} \Delta x_{\nu} \Delta x_{\rho} \Delta x_{\sigma}}{\Delta
x^4} \Biggr\} \frac{ \Gamma( \frac{D}2 \!-\! 1)}{ 4 \pi^{\frac{D}2}
\Delta x^{D-2}} \; . \qquad
\end{eqnarray}

Let us see what happens if we exploit what the mathematical
physicists assert to be the freedom to add a constant to equation
(\ref{flatdisp}),
\begin{equation}
\partial^{10} S_{\rm flt}(x \!-\! x') = 8 \Bigl( \frac{D \!-\! 2}{D
\!-\! 3}\Bigr)^2 i\delta^D(x \!-\! x') + M^D \; . \label{flatmath}
\end{equation}
At this point some mathematical physicists object that the only 
dimensionful constant on de Sitter is $H$, so any constant we add
to the propagator equation must be proportional to $H^{D}$, which 
vanishes in the flat space limit. This is sophistry. Morrison's 
argument is based on the vanishing of expression (\ref{constant}) 
so it applies to an {\it arbitrary} constant. If the ambiguity 
is real then we must be able to add any constant to the propagator 
equation, including one which fails to vanish in the flat space 
limit.

The result of changing the flat space propagator equation to 
(\ref{flatmath}) is to change the structure function by a term we 
might call $\Delta S(x - x')$ which obeys,
\begin{equation}
\partial^4 \Delta S(x \!-\! x') = \frac{ M^D \Delta x^6}{48 D (D
\!+\! 2) (D \!+\! 4)} \; .
\end{equation}
The resulting change in the spin two part of the propagator is,
\begin{eqnarray}
\lefteqn{ {\bf T}_{\mu\nu\rho\sigma} \Delta S = \frac{(D \!+\! 1) (D
\!-\! 3)^2 M^D}{8 (D \!+\! 4) (D \!+\! 2) D (D \!-\! 1) (D \!-\!
2)^2} \Biggl\{ (D^2 \!+\! 2 D \!-\! 4) \eta_{\mu (\rho}
\eta_{\sigma) \nu} \Delta x^2 } \nonumber \\
& & \hspace{-.7cm} - (D\!+\! 2) \eta_{\mu\nu} \eta_{\rho\sigma}
\Delta x^2 \!-\! 4 D \Delta x_{(\mu} \eta_{\nu) (\rho} \Delta
x_{\sigma)} \!+\! 4 \Bigl[ \Delta x_{\mu} \Delta x_{\nu}
\eta_{\rho\sigma} \!+\! \eta_{\mu\nu} \Delta x_{\rho} \Delta
x_{\sigma}\Bigr] \Biggr\} . \qquad \label{flatshift}
\end{eqnarray}
It is difficult to understand what sort of state could give rise to
the long range correlations evident in expression (\ref{flatshift}).

The addition of ill-behaved terms such as (\ref{flatshift}) is
typical when one solves the propagator equation (\ref{flateqn1})
without demanding that the solution be a propagator. The structure
function must be a mode sum in order to give a true propagator,
which precludes the addition of constants, or any other function
annihilated by the projectors. This is obvious in the spatial
Fourier basis appropriate to flat space, and to de Sitter in open
coordinates. The case for an extra constant seems better for de
Sitter in closed coordinates, because the constant can be
represented as $Y_{000}(\chi,\theta,\phi) \times
Y^*_{000}(\chi',\theta',\phi')$. However, it will be noted that the
temporal dependence doesn't quite work out. There are two linearly
independent zero modes, only one of which is constant, and a proper
mode sum must involve both of them.

For those mathematical physicists who still insist on $M \sim H$ we
should note that it is not necessary to take the flat space limit to
see that adding the constant is problematic. The fact that graviton 
modes in transverse-traceless-spatial gauge agree with those of the 
MMC scalar \cite{Grishchuk}, and the analysis of Ford and Parker for 
the latter \cite{FP}, imply that transverse-traceless-spatial 
gravitons suffer from infrared problems for all FRW geometries whose 
first slow roll parameter $\epsilon \equiv -\dot{H}/H^2$ is constant 
and in the range $0 \leq \epsilon \leq 2(D-1)/D$. As we have already 
pointed out, the spin two sector of the graviton propagator for all 
these cases can be represented as (\ref{Spin2}), with only a slight 
generalization of the transverse-traceless projector ${\bf P}_{\mu
\nu}^{~~ \rho\sigma}$. (Indeed, the ``generalization'' consists of 
undoing the specialization of the original operator to de Sitter 
\cite{SPW}.) In particular, the propagator equation for all these 
cases would allow the addition of terms annihilated by ${\bf P}_{
\mu\nu}^{~~\rho\sigma}$, and the consequent additions to the 
graviton propagator would be as unphysical as the one 
(\ref{flatshift}) we found for flat space. Only for the de Sitter 
case of $\epsilon = 0$ does the siren call of additional symmetry 
beguile the mathematically inclined to dispute the passage from 
(\ref{almost}) to (\ref{disputed}).

Consideration of the photon propagator on de Sitter background makes
the argument even stronger. The spin one sector of the photon can be
given a representation comparable to (\ref{Spin2}), for which it was
indeed the paradigm \cite{MTW3},
\begin{equation}
i \Bigl[\mbox{}_{\mu} \Delta^1_{\rho}\Bigr](x;x') = -\frac1{2 H^2}
{\bf P}_{\mu}^{~ \nu}(x) \times {\bf P}_{\rho}^{~ \sigma}(x') \Bigl[
\mathcal{R}_{\nu\sigma}(x;x') \mathcal{S}_T(x;x')\Bigr] \; .
\label{Spin1}
\end{equation}
The transverse projector ${\bf P}_{\mu}^{~ \nu}$ in (\ref{Spin1}) is
constructed from the field strength tensor the same way as ${\bf
P}_{\mu\nu}^{~~ \alpha\beta}$ was constructed from the Weyl tensor
\cite{SPW},
\begin{equation}
F^{\alpha\beta} = \mathcal{P}_{\mu}^{~\alpha\beta} \times A^{\mu}
\qquad \Longrightarrow \qquad {\bf P}_{\mu}^{~ \nu} \equiv
\mathcal{P}_{\mu}^{~ \nu\alpha} D_{\alpha} \; .
\end{equation}
The transverse projectors annihilate constants in the spin one
sector the same way that the transverse-traceless projectors do
(\ref{constant}) in the spin two sector,
\begin{equation}
{\bf P}_{\mu}^{~\nu}(x) \times {\bf P}_{\rho}^{~ \sigma}(x') \Bigl[
\mathcal{R}_{\nu \sigma}(x;x') \Bigr] = 0 \; . \label{constant2}
\end{equation}
So there is equal justification for adding a constant to the
equation for the spin one structure function \cite{MTW3},
\begin{equation}
\Bigl[ \square \!-\! (D \!-\! 2) H^2\Bigr]^2 \Bigl[ \square' \!-\!
(D \!-\! 2) H^2\Bigr] \mathcal{S}_T(x;x') = -2 H^2 \, \frac{i
\delta^D(x \!-\! x')}{\sqrt{-g}} \; . \label{righteqn}
\end{equation}
The only problem is: {\it equation (\ref{righteqn}) already gives a
de Sitter invariant propagator \cite{TW7} which mathematical
physicists accept} \cite{Higuchi2}. Adding any nonzero constant to
(\ref{righteqn}) would produce a different, and incorrect result.
The freedom Morrison claims to have discovered is simply not
present.

\subsection{Inequivalence of the two propagators}\label{not}

The coincidence limit provides a very simple way of seeing that no
de Sitter invariant solution to the propagator equation
(\ref{propeqn}) can be physically equivalent to our de Sitter
breaking propagator. The coincidence limit of our result is
\cite{KMW},
\begin{eqnarray}
\lefteqn{ \lim_{x' \rightarrow x} i\Bigl[\mbox{}_{\mu\nu}
\Delta^2_{\rho\sigma}\Bigr](x;x') = \Bigl( {\rm Const} \Bigr) \Bigl[
2 g_{\mu (\rho} g_{\sigma ) \nu} \!-\! \frac2{D} \, g_{\mu\nu}
g_{\rho\sigma}\Bigr] } \nonumber \\
& & \hspace{1.5cm} + \Biggl( \frac{H^{D-2}}{ (4\pi)^{\frac{D}2}}
\frac{\Gamma(D \!-\! 1)}{\Gamma(\frac{D}2)} \, 2 H t \!+\! {\rm
Const}\Biggr) \Bigl[2 g^{\perp}_{\mu (\rho} g^{\perp}_{\sigma) \nu}
\!-\! \frac2{D \!-\! 1} \, g^{\perp}_{\mu\nu} g^{\perp}_{\rho\sigma}
\Bigr] \; , \qquad \label{coincidence}
\end{eqnarray}
where $g^{\perp}_{\mu\nu}$ is the purely spatial part of the metric.
By contrast, the coincidence limit of a de Sitter invariant solution
to the propagator equation could only have the de Sitter invariant
first line of (\ref{coincidence}); it could never contain the
explicitly time dependent factor of $H t$, or the de Sitter breaking
tensor structure of the second line. These de Sitter breaking
features agree with the traceless part of the noncovariant graviton
propagator \cite{TW1}, and with the result in
transverse-traceless-spatial gauge \cite{MTW4}. The physical origin
of the secular growth evident in (\ref{coincidence}) is the same as
for the coincidence limit of the MMC scalar \cite{VFLS}: as time
progresses, more and more modes experience first horizon crossing
and become constant. This is not a gauge artifact but rather the
mechanism by which quantum fluctuations from primordial inflation
become fossilized so that they can be observed at late times.

\begin{figure}
\hskip -3.5cm \includegraphics[width=3.8cm,height=3.0cm]{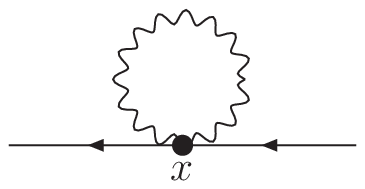}
\hskip 1cm \includegraphics[width=3.8cm,height=3.0cm]{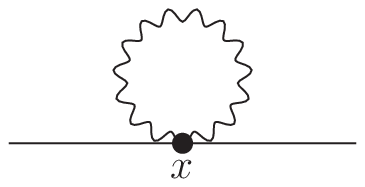}
\hskip 1cm \includegraphics[width=3.8cm,height=3.0cm]{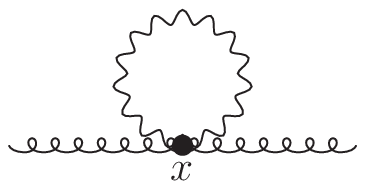}
\caption{\tiny Coincident graviton propagator contributions to
various matter field 1PI 2-point functions. The leftmost diagram is
from the one loop fermion self-energy \cite{MW1,SPM}; the center
figure is from the one loop scalar self-mass-squared \cite{KW1}; and
the rightmost diagram is from the one loop vacuum polarization
\cite{LW}.} \label{matgraphs}
\end{figure}

\begin{figure}
\hskip -3cm \includegraphics[width=4.0cm,height=3.0cm]{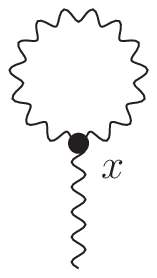}
\includegraphics[width=4.0cm,height=3.0cm]{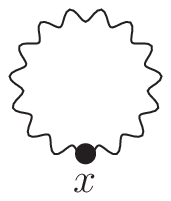}
\includegraphics[width=4.0cm,height=3.0cm]{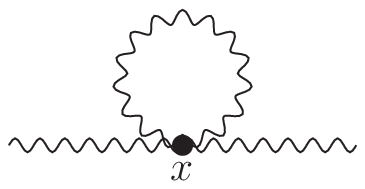} \vskip 1cm
\caption{\tiny Coincident graviton propagator contributions to
various one loop 1PI functions and expectation values in pure
gravity. The leftmost diagram is from the one loop graviton 1-point
function \cite{TW2}; the center figure is from the one loop
expectation value of the square of the Weyl tensor \cite{PMW,PMTW2};
and the rightmost diagram is from an old computation of the graviton
self-energy with a momentum cutoff \cite{TW8} which is being re-done
with dimensional regularization.} \label{QGRgraphs}
\end{figure}

A mathematical physicist might be tempted to dismiss the coincidence
limit of a propagator as too singular to provide a good comparison
but it makes perfect sense in dimensional regularization.
Figures~\ref{matgraphs} and \ref{QGRgraphs} also show that the
coincident graviton propagator contributes to every single one of
the graviton loops for which fully dimensionally regulated results
have so far been obtained on de Sitter background
\cite{TW2,MW1,KW1,PMW,SPM,LW}. And the de Sitter breaking evident in
the coincidence limit (\ref{coincidence}) is of course present as
well for ${x'}^{\mu} \neq x^{\mu}$. Taking the coincidence limit is
just the most obvious way of demonstrating that de Sitter breaking
is a real effect.

Morrison argues that our de Sitter breaking propagator is
nonetheless ``physically equivalent'' to his de Sitter invariant
solution to the propagator equation. The argument consists of
showing that smearing with the transverse-traceless test functions
of Fewster and Hunt \cite{FH} makes the de Sitter breaking
difference drop out \cite{IAM},
\begin{equation}
\int \!\! d^Dx \, f^{\mu\nu}_1(x) \! \int \!\! d^Dx' \,
f^{\rho\sigma}_2(x') \times i\Bigl[ \mbox{}_{\mu\nu} \Delta^{\rm
2br}_{\rho\sigma} \Bigr](x;x') = 0 \; . \label{smear}
\end{equation}
Morrison interprets (\ref{smear}) to mean that the de Sitter
breaking contributions to our propagator are pure gauge. That cannot
be so because the identical time dependence occurs in the completely
fixed transverse-traceless-spatial gauge \cite{MTW4}. The actual
explanation is that transverse-traceless smearing test functions do
not completely scrutinize free graviton fields because no sequence
of them can be made to approach a delta function. In this feature
they show a critical difference from the scalar test functions
\cite{Jaffe} on which they were surely based.

To see the problem we may as well consider $D=4$ flat space.
Mathematical physicists working in constructive quantum field theory
consider a scalar field $\varphi(x)$ to be an operator-valued
distribution which is too singular to be studied in its original
form, as a function of spacetime \cite{Jaffe}. They instead
``smear'' $\varphi(x)$ against smooth test functions $f(x)$,
\begin{equation}
\varphi(x) \longrightarrow \varphi[f] \equiv \int \!\! d^4x \, f(x)
\varphi(x) \; .
\end{equation}
The point is to be able to prove nonperturbative theorems. Of course
there is no formalism of quantum gravity which makes sense beyond
the realm of regulated perturbation theory, and there is absolutely
no need for smearing when using regulated perturbation theory as we
are. However, nothing is lost by using the smearing formalism for
scalars because one can form delta sequences of test functions which
approach a delta function,
\begin{equation}
f_n(x;x') \longrightarrow \delta^4(x \!-\! x') \; .
\end{equation}
Were we to represent the test function in Fourier space the
analogous statement would be,
\begin{equation}
f_n(t,\vec{x}) = \int \!\! \frac{d^3k}{(2\pi)^3} \, e^{i \vec{k}
\cdot \vec{x}} \widetilde{f}_n(t,\vec{k}) \qquad \Longrightarrow
\qquad \widetilde{f}_n(t,\vec{k}) \longrightarrow \delta(t \!-\! t')
e^{-i \vec{k} \cdot \vec{x}'} \; . \label{scaltest}
\end{equation}

Let us now consider how smearing works for linearized gravitons. A
general transverse-traceless test function might be defined as,
\begin{equation}
f^{\mu\nu}(t,\vec{x}) = \int \!\! \frac{d^3k}{(2\pi)^3} \, e^{i
\vec{k} \cdot \vec{x}} \sum_{\lambda = \pm}
\epsilon^{\mu\nu}(\vec{k},2\lambda) \,
\widetilde{f}(t,\vec{k},\lambda) \; .
\end{equation}
The graviton polarization tensors
$\epsilon^{\mu\nu}(\vec{k},2\lambda)$ can be expressed by taking
products of photon polarization vectors,
\begin{equation}
\epsilon^{\mu\nu}(\vec{k},2\lambda) =
\epsilon^{\mu}(\vec{k},\lambda) \times
\epsilon^{\nu}(\vec{k},\lambda) \; .
\end{equation}
We define the latter to be purely spatial and transverse.
Transversality can be explicitly enforced by expressing the Fourier
wave vector in spherical coordinates $\vec{k} = k \widehat{r}$ with
the spherical unit vectors defined as usual,
\begin{eqnarray}
\widehat{r} & \equiv & \sin(\theta) \cos(\phi) \widehat{x} \!+\!
\sin(\theta) \sin(\phi) \widehat{y} \!+\! \cos(\theta) \widehat{z}
\; , \\
\widehat{\theta} & \equiv & \cos(\theta) \cos(\phi) \widehat{x}
\!+\! \cos(\theta) \sin(\phi) \widehat{y} \!-\! \sin(\theta)
\widehat{z} \; , \\
\widehat{\phi} & \equiv & -\sin(\phi) \widehat{x} \!+\! \cos(\phi)
\widehat{y} \; .
\end{eqnarray}
Because $\lambda = \pm 1$ we can define a general transverse
polarization vector as,
\begin{equation}
\epsilon^i(\vec{k},\lambda) \equiv \frac1{\sqrt{2}} \Bigl(
\widehat{\theta}^i + i \lambda \widehat{\phi}^i \Bigr) \qquad
\Longrightarrow \qquad \eta_{\mu\nu} \epsilon^{\mu}(\vec{k},\lambda)
\epsilon^{\nu*}(\vec{k},\lambda') = \delta_{\lambda \lambda'} \; .
\end{equation}

The rest of the derivation is straightforward. If any sequence of
$\widetilde{f}(t,\vec{k},\lambda)$ did lead to a delta function
comparison with (\ref{scaltest}) suggests that it would be,
\begin{equation}
\widetilde{f}_n(t,\vec{k},2\lambda) \longrightarrow \delta(t \!-\!
t') e^{-i \vec{k} \cdot \vec{x}'} \Bigl( \delta_{\lambda +} \!+\!
\delta_{\lambda -}\Bigr) \; . \label{ansatz}
\end{equation}
To see that the Fourier transform of (\ref{ansatz}) cannot produce a
4-dimensional delta function, simply choose the $\widehat{z}$ axis
of $\vec{k}$ parallel to $\Delta \vec{x} \equiv \vec{x} - \vec{x}'$,
which leaves only the polarization tensors depending upon the
azimuthal angle $\phi$. For either polarization one finds,
\begin{equation}
\int_0^{2\pi} \!\! d\phi \, \epsilon^i \epsilon^j = \frac{\pi}2
\sin^2(\theta) \Bigl(3 \, \frac{\Delta x^i \Delta x^j}{\Delta x^2} -
\delta^{ij} \Bigr) \; .
\end{equation}
Performing the $\theta$ and $r$ integrations gives,
\begin{equation}
f^{ij}_n(t,\vec{x}) \longrightarrow \frac{\delta(t \!-\! t')}{8 \pi
\Delta x^3} \Bigl( 3 \, \frac{\Delta x^i \Delta x^j}{\Delta x^2} -
\delta^{ij} \Bigr) \; . \label{best}
\end{equation}
Expression (\ref{best}) is transverse, traceless and purely spatial,
but it doesn't let us recover the original graviton field. In fact,
expression (\ref{best}) vanishes at $\vec{x}' = \vec{x}$ if one
employs it inside an integral for which the angular average gives $3
\Delta x^i \Delta x^j/\Delta x^2 \longrightarrow \delta^{ij}$! In
particular, one can never approach the coincidence limit by smearing
two transverse-traceless test functions as in Morrison's identity
(\ref{smear}). So it is not correct to say that our de Sitter
breaking propagator is physically equivalent to his de Sitter
invariant solution of the propagator equation; rather he has not
permitted himself to scrutinize the difference between them with
sufficient resolution. And we might note that it was already obvious
from the agreement of the linearized Weyl-Weyl correlators
\cite{PMW} that a high resolution probe is needed to detect the
difference.

\subsection{Tensor power spectrum gives the coincidence
limit}\label{coinc}

A probe with the required sensitivity is at hand in the form of the
primordial tensor power spectrum. The tensor power spectrum derives
from the late time limit of the spatial Fourier transform, in open
coordinates, of the 2-point correlator of the graviton field in
transverse-traceless-spatial gauge \cite{reviews}. In our notation
(\ref{hdef}) everything but the late time limit would be,
\begin{eqnarray}
\Delta^2_h(k,t) & \equiv & \frac{k^3}{2 \pi^2} \! \int \!\!
d^3x \, e^{-i \vec{k} \cdot \vec{x}} \Bigl\langle
\Omega \Bigl\vert \frac{\sqrt{2} \, \kappa}{a^2(t)} \,
h_{ij}(t,\vec{x}) \times \frac{\sqrt{2} \, \kappa}{a^2(t)} \,
h_{ij}(t,\vec{0}) \Bigr\vert \Omega \Bigr\rangle \; , \label{power1}
\qquad \\
& = & \frac{k^3}{2 \pi^2} \times 32\pi G \times 2 \times \vert
u(t,k) \vert^2 \; . \qquad \label{power2}
\end{eqnarray}
The actual tensor power spectrum is defined by evolving the tensor
mode function $u(t,k)$ past the 1st horizon crossing time (which is
$t_k = H^{-1} \ln(k/H)$ for de Sitter) at which they ``freeze in''
to $u(t,k) \sim H/\sqrt{2 k^3}$. The result for de Sitter is,
\begin{equation}
\lim_{t \gg t_k} \Delta^2_h(k) \longrightarrow \frac{16}{\pi} \, G
H^2 \; . \label{truepower}
\end{equation}
The absence of any dependence on the wave number $k$ is known as
``scale invariance''. It is these two features of freezing in and
scale invariance which enable us to observe the primordial power
spectra.

Although the ``tensor power spectrum'' is defined as the late time
limit of expressions (\ref{power1}-\ref{power2}), it is better to
retain the time dependent formulae for our current discussion. The
point of this subsection is that there is a simple relation between
the power spectrum and the trace of the coincident spin two
propagator we have been debating,
\begin{eqnarray}
16 \pi G g^{\mu\rho}(x) g^{\nu\sigma}(x) \!\times\! i
\Bigl[\mbox{}_{\mu\nu} \Delta^2_{\rho\sigma}\Bigr](x;x) &
\!\!\!=\!\!\! & \frac52 \!\times \! \int \!\! \frac{d^3k}{(2\pi)^3}
\, 32\pi G \!\times\! 2 \!\times\! \vert u(t,k)\vert^2 \; , \qquad
\label{IRdiv1} \\
& \!\!\!=\!\!\! & \frac52 \!\times\! \int \!\! \frac{d^3k}{(2\pi)^3}
\, \frac{2 \pi^2}{k^3} \, \Delta^2_h(k,t) \; , \\
& \!\!\!=\!\!\! & \frac52 \!\times\! \int \! \frac{dk}{k} \,
\Delta^2_h(k,t) \; . \label{IRdiv}
\end{eqnarray}
The factor of $\frac52$ derives from the contribution of three
constrained fields to the gauge-fixed but unconstrained propagator
$i[\mbox{}_{\mu\nu} \Delta^2_{\rho\sigma}](x;x')$, whereas the
tensor power spectrum has only the two dynamical gravitons.

Because the tensor power spectrum is a gauge invariant observable,
relation (\ref{IRdiv}) provides an enormously powerful insight into
the de Sitter breaking time dependence of the coincidence limit
(\ref{coincidence}) of the spin two sector of the graviton
propagator. First, we note that the naive mode sum is infrared
divergent. The physical origin of this infrared divergence is the
same as the analogous scalar infrared divergence which was discussed
at the end of subsection~\ref{mustnot}. With either of the two
standard fixes \cite{AV,TW6} the naive mode sum is effectively cut
off at some fixed lower limit corresponding to the co-moving wave
number of the longest wave length which is initially in Bunch-Davies
vacuum. The time dependence of the result (\ref{IRdiv}) arises
because the time dependent power spectrum $\Delta^2_h(k,t)$ assumes
its asymptotic form (\ref{truepower}) at the time of first horizon
crossing, which is $t_k \sim H^{-1} \ln(k/H)$ for de Sitter. So the
integral becomes,
\begin{equation}
\frac52 \!\times\! \int_{H}^{He^{Ht}} \frac{dk}{k} \times
\frac{16}{\pi} \, G H^2 = \frac{40}{\pi} \, G H^2 \times H t \; .
\label{timedep}
\end{equation}
Substituting relation (\ref{coincidence}) to the left hand side of
(\ref{IRdiv1}) gives complete agreement with (\ref{timedep}). Note
again the complete impossibility of accommodating a de Sitter
invariant solution to the propagator equation.

A closely related point has been made before in the context of the
totally gauge fixed and constrained propagator in
transverse-traceless-spatial gauge. An on-shell field redefinition
which carries this de Sitter breaking propagator to a de Sitter
invariant one has been given in \cite{HMM}. Of course it is not
possible to change the propagator, while preserving the gauge-fixed
and constrained field equations, without altering the canonical
commutation relations \cite{MTW4}, so their construction is really
an excursion into non-canonical quantization. One consequence of the
altered quantization scheme is that the usual definition of the
tensor power spectrum produces a result which breaks scale
invariance \cite{MTW4}. Mathematical physicists retort that one must
employ a new, ``gauge invariant'' definition of the tensor power
spectrum which recovers the usual, scale invariant result
\cite{HMM}. They have so far neglected to specify this definition
but one might observe first, that any quantity becomes invariant
when defined in a unique gauge such as transverse-traceless-spatial
gauge \cite{TW9}, and second, that any revised definition of the
power spectrum which amounts to using the old, de Sitter breaking
propagator in the old way is indistinguishable from simply conceding
that free gravitons break de Sitter invariance.

We close by anticipating an objection which might be raised against
appealing to the observability of the tensor power spectrum in the
context of de Sitter results. The argument goes that perfect de 
Sitter inflation never ends, therefore modes which have experienced 
first horizon crossing will never re-enter the horizon, which is 
necessary for them to produce a detectable spatial variation. Hence
the power spectrum of de Sitter is unobservable and it cannot be 
invoked to prove de Sitter breaking. We ask those who attempt to 
escape the inevitability of de Sitter breaking through recourse to 
this argument to consider a multi-scalar inflation model in which 
the usual decline of $H(t)$ ceases for a period of time which is 
controlled by a ``clock'' provided by one of the other scalars. 
During the $\dot{H}(t) = 0$ phase the geometry is locally de 
Sitter and modes which experience first horizon crossing should have 
the scale invariant amplitude. However, inflation eventually ends so 
these modes can experience second horizon crossing and become 
observable to a late time observer. Would such a late time observer 
measure their power spectrum to be scale invariant? If the answer is 
conceded to be ``yes'' then it must be admitted that the coincidence 
limit of the graviton propagator shows de Sitter breaking.

\section{Discussion}\label{discuss}

The recent paper by Morrison \cite{IAM} demonstrates the remarkable
convergence of opinion on the graviton propagator which has taken
place over the past few years. In particular, the allowed gauges are
universally agreed, as is almost all of the spacetime dependence and
tensor structure in any allowed gauge. The remaining points of
disagreement have been narrowed to just seven issues, which we
summarize from our perspective:
\begin{enumerate}
\item{It is no more valid to define tachyonic mode sums by
analytic continuation in the scalar mass-squared than to
analytically continue in the dimension, in the signature or in the
deceleration parameter. Demonstrating that these analytic
continuations all give the same result only shows that they all make
the same error of incorporating negative norm states.}
\item{The massive scalar propagator breaks de Sitter invariance for
all $M^2 \leq 0$ and, by continuity, de Sitter breaking shows up even 
for $M^2 > 0$ in the solution which is truly an analytic function of 
$M^2$. (See also \cite{Anderson:2013ila,Anderson:2013zia}.)
Denying this leads to the nonsensical conclusion that a tachyonic
scalar with $M^2 = -N (N + D-1) H^2$ decays, but making $M^2$
slightly {\it more} tachyonic stabilizes it.}
\item{The de Sitter breaking of the massless, minimally coupled
scalar propagator is not due to its isolated zero mode but rather to
the fact that all its mode functions approach scale invariant
constants. Graviton mode functions approach the same scale invariant
constants. These are physical effects, not gauge artifacts, and they
show up in closed coordinates as well as on the cosmological patch.
This behavior is why the scalar and tensor power spectra from
primordial inflation can be observed during the current epoch.}
\item{There is no ambiguity in the equation for the spin two
structure function if one requires that the propagator and the
projection operator be positive norm mode sums. Violating this 
precept in flat space would compromise unitarity.}
\item{Because no sequence of the transverse-traceless smearing
functions proposed by Fewster and Hunt \cite{FH} recovers the
point-wise graviton field, equality of two smeared propagators does
not imply their physical equivalence.}
\item{The coincidence limit of the graviton propagator --- which
shows up in every fully dimensionally regulated graviton loop that
has so far been computed \cite{TW2,MW1,KW1,PMW,SPM,LW} --- reveals
that our de Sitter breaking propagator is not physically equivalent
to any de Sitter invariant solution to the propagator equation.}
\item{The time independence and scale invariance of the tensor
power spectrum require that the graviton propagator breaks de Sitter
invariance.}
\end{enumerate}
Rather than regarding the continuing debate over these points as a
distasteful controversy to be deplored and avoided, we view it as
the embodiment of the scientific method. We hope this paper will
continue the process, and we foresee complete concurrence in the
near future.

Morrison has carefully laid out the procedures necessary to extract
a de Sitter invariant solution from the graviton propagator
equation. These are:
\begin{itemize}
\item{One must add a constant to the equation which defines the
structure function of the spin two projection operator; and}
\item{One must consider the scalar propagator to be both de Sitter
invariant and a meromorphic function of the scalar mass-squared,
with simple poles at $M^2 = -N (N + D -1) H^2$.}
\end{itemize}
Morrison has also derived the precise difference between our de
Sitter breaking structure functions and the de Sitter invariant
structure functions that result from following his procedures. These
differences are rather small for the spin two sector --- which had
to be the case in view of the fact that the Weyl-Weyl correlators
agree \cite{PMW,PMTW2} --- but they have the significant effect of
making the coincident propagator time dependent. And we emphasize
that the coincident propagator enters every single one of the
dimensionally regulated graviton loops which have so far been
computed \cite{TW2,MW1,KW1,PMW,SPM,LW}. It is obvious the two
solutions to the propagator equation mediate different physics, and
it is important to resolve which one is the true propagator.

\centerline{\bf Acknowledgements}

We are grateful for conversations on this subject with S. Deser, E.
Kiritsis and T. Prokopec. This work was partially supported by NWO
Veni Project \# 680-47-406, by European Union program Thalis
ESF/NSRF 2007-2013 MIS-375734, by European Union (European Social
Fund, ESF) and Greek national funds through the Operational Program
``Education and Lifelong Learning'' of the National Strategic
Reference Framework (NSRF) under ``Funding of proposals that have
received a positive evaluation in the 3rd and 4th Call of the ERC
Grant Schemes'', by NSF grant PHY-1205591, and by the Institute for
Fundamental Theory at the University of Florida.

\end{document}